%% file: 00-main.tex
\documentclass[a4paper,12pt]{article}
\usepackage[ansinew]{inputenc}
\usepackage{amsthm,amssymb,amsmath,amsfonts,amstext}
\usepackage[english]{babel}
\usepackage{array,latexsym}
\usepackage{graphicx,lscape,amsmath,amssymb}

\newcommand{\R}{I\!\!R}

\usepackage{commath,mathrsfs,float,bbm,dsfont,soul,subfigure,tikz}
\usepackage{setspace,newlfont,authblk,lscape,longtable,multirow,comment,xfrac,indentfirst}
\usepackage[T1]{fontenc}
\usepackage[round]{natbib}
\usepackage[colorlinks=true, linkcolor=blue, citecolor=blue, filecolor=blue, urlcolor=blue, breaklinks=true] {hyperref}
\usepackage[left=2.5cm,right=2.5cm,top=2.5cm,bottom=2.5cm]{geometry}

\newcolumntype{G}{> {\columncolor[gray]{0.8}}c}

\title{Modelling the Extremes of Seasonal Viruses and Hospital Congestion: The Example of Flu in a Swiss Hospital}

\author{
Setareh Ranjbar$^{(1)}$, Eva Cantoni$^{(2)}$, Val\'erie Chavez-Demoulin$^{(1)}$, Giampiero Marra$^{(3)}$, Rosalba Radice$^{(4)}$, Katia Jaton-Ogay$^{(5)}$\\
 \small $^{(1)}$ Faculty of Business and Economics, University of Lausanne,  \\
$^{(2)}$ Research Center for Statistics, GSEM, University of Geneva
\\
$^{(3)}$ Department of Statistical Science,  University College London
\\
$^{(4)}$ Faculty of Actuarial Science and Insurance, Cass Business School
\\
$^{(5)}$ Institut Universitaire de Microbiologie, CHUV, Lausanne
} 

\date{}

\begin{document}

\maketitle
\begin{abstract}
Viruses causing flu or milder coronavirus colds are often referred to as ``seasonal viruses'' as they tend to subside in warmer months. In other words, meteorological conditions tend to impact the activity of viruses, and this information can be exploited for the operational management of hospitals. 
In this study, we use three years of daily data from one of the biggest hospitals in Switzerland and focus on modelling the extremes of hospital visits from patients showing flu-like symptoms and the number of positive cases of flu. We propose employing a discrete Generalized Pareto distribution for the number of positive and negative cases, and a Generalized Pareto distribution for the odds of positive cases. Our modelling framework allows for the parameters of these distributions to be linked to covariate effects, and for outlying observations to be dealt with via a robust estimation approach. Because meteorological conditions may vary over time, we use meteorological and not calendar variations to explain hospital charge extremes, and our empirical findings highlight their significance. We propose a measure of hospital congestion and a related tool to estimate the resulting CaRe (Charge-at-Risk-estimation) under different meteorological conditions. The relevant numerical computations can be easily carried out using the freely available \texttt{GJRM R} package. The introduced approach could be applied to several types of seasonal disease data such as those derived from the new virus SARS-CoV-2 and its COVID-19 disease which is at the moment wreaking havoc worldwide. The empirical effectiveness of the proposed method is assessed through a simulation study.  

\vspace*{1cm}

\end{abstract}

\newpage
\section{Introduction}
\input{01-Intro.ch}
\section{Robust Regression Methodology for Peaks-Over-Threshold}\label{sct:method}
\input{02-Method.ch}

\section{Modelling Flu Extremes}
\input{03-Models.ch}

\section{Simulation Study}
\input{05-simulation.ch}

\section{Conclusion}
\input{05-conclusion.ch}

\section*{Acknowledgments}
The simulations were performed at the University of Geneva using the Baobab cluster. The research was partially funded by the Swiss National Science Foundation SNF (first and third authors).

\bibliographystyle{elsarticle-harv}
\bibliography{mybib}

\clearpage
\appendix

\setcounter{page}{1}
\setcounter{section}{0}

\newcommand{\ssection}[1]{%
\section[#1]{\centering\normalfont\scshape #1}}

\renewcommand{\thesection}{\Alph{section}}
\renewcommand{\thesubsection}{\Alph{section}.\arabic{subsection}}

\setcounter{table}{0}
\setcounter{figure}{0}
\setcounter{equation}{0}
\renewcommand\thetable{S\arabic{table}}  
\renewcommand\thefigure{S\arabic{figure}}  
\renewcommand\theequation{S\arabic{equation}}  

\newpage
\begin{center}
 \Large{ Supplementary Materials for \\[10pt] 
\textit{Modelling the Extremes of Seasonal Viruses and Hospital Congestion: The Example of Flu in a Swiss Hospital} \\[10pt]  by Setareh Ranjbar, Eva Cantoni, Val\'erie Chavez-Demoulin, Giampiero Marra, Rosalba Radice, Katia Jaton-Ogay
}
\end{center}
\input{06-appendix.ch}
\input{07-appendixSoft.ch}\label{sct:appexSoft}
\input{Supp.ch}\label{sct:appexSupp}
\input{08-simulation-figures.ch}\label{sct:appexFig}

\end{document}

%% file: 01-Intro.ch
Congestion is of paramount concern for most large size hospitals. At the time of writing, hospitals across the world are experiencing congestion due to the coronavirus pandemic. Predicting both the number of visits to the hospital and potential pandemic considerably helps the operational management of hospitals. Empirical evidence suggests that the flu virus is more vulnerable in warm weather, hence making it more common for epidemics to proliferate in fall and winter months in the northern hemisphere. The role of weather in the spread of flu is not yet fully understood and researchers have attempted to address this question. \cite{roussel2016} studied the impact of six climate variables (related to temperature, humidity and sunshine) on flu spread, whereas \cite{towers2013} analysed waves of influenza and climate patterns. \cite{davis2012} examined the hypothesis that cold and/or dry weather enhances human pneumonia and influenza mortality, whereas \cite{firestone2012} quantified the association between the hazard of flu infection and air temperature, humidity, rainfall and wind velocity. It has been generally found that flu's transmission is mostly dependent on humidity and temperature \citep{lowen2014}, with cold and dry weather making flu more active. In this paper, we approach the problem from the point of view of hospitals facing the risk of congestion and hence the need for assessing the efficiency of the flu testing process. Specifically, we aim at understanding and quantifying the impact of weather related variables on the probability of obtaining: a high number of positive flu tested patients (epidemic), a high number of negative flu tested patients (inefficiency), and high values of positive odds (number of positive flu over negative flu tested patients). 

We use three years (2016/2017, 2017/2018 and 2018/2019) of daily data from 01/07/2016 to 21/06/2019 which give us $n=1086$ observations on the number of visits and positive cases for flu at the Lausanne University Hospital\footnote{https://www.lausanneuniversityhospital.com/home} (CHUV), one of the largest hospitals in Switzerland, with a capacity of 1000 somatic beds. 

\begin{figure}[H]
    \centering
    \includegraphics[width=0.8\linewidth]{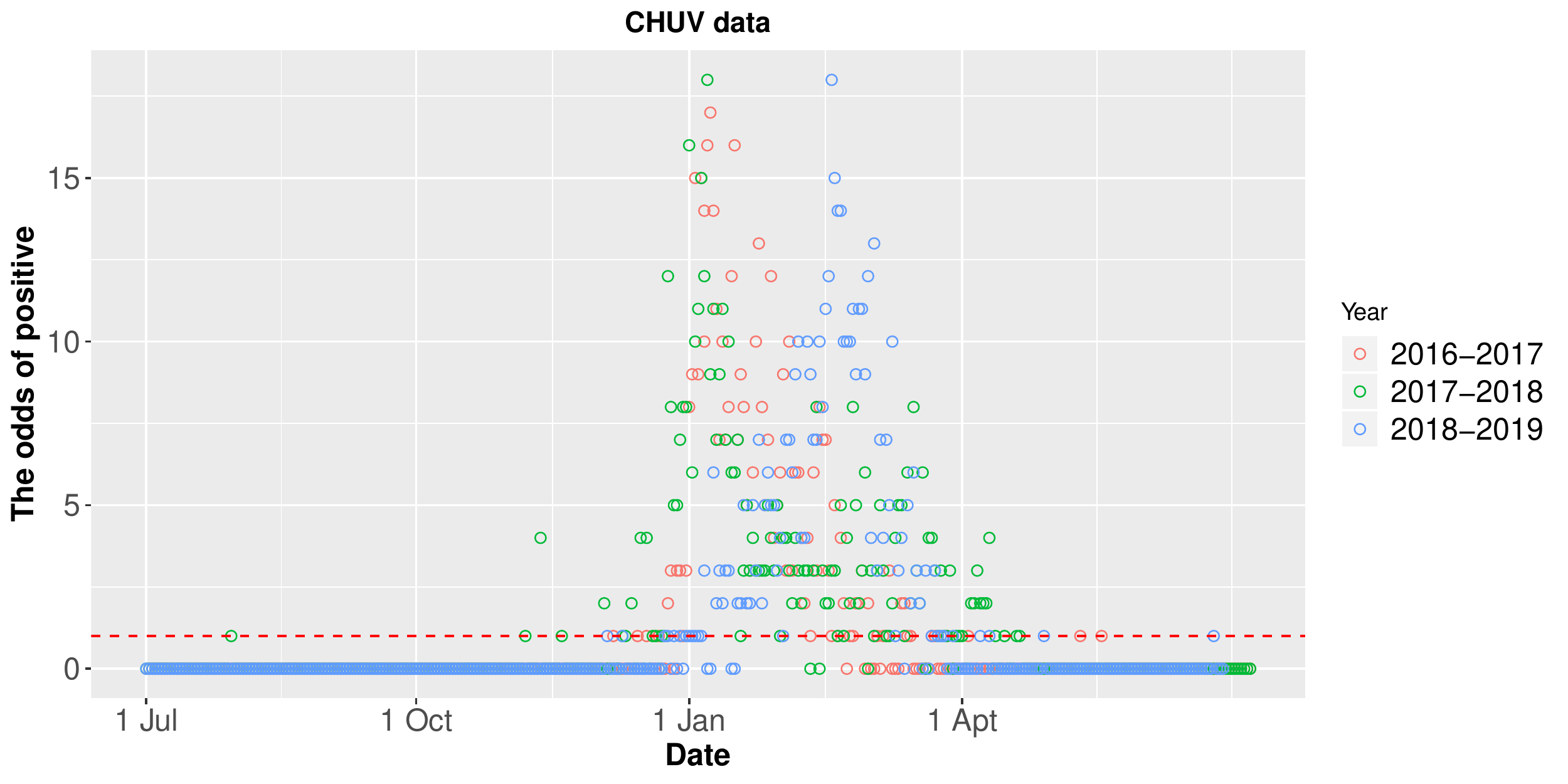}
    \caption{Daily number of flu positive cases tested patients from 01/07/2016 to 21/06/2019. The red line shows the threshold defining exceedances.}
        \label{fig:raw-positive}
\end{figure}

\begin{figure}[H]
    \centering
    \includegraphics[width=0.8\linewidth]{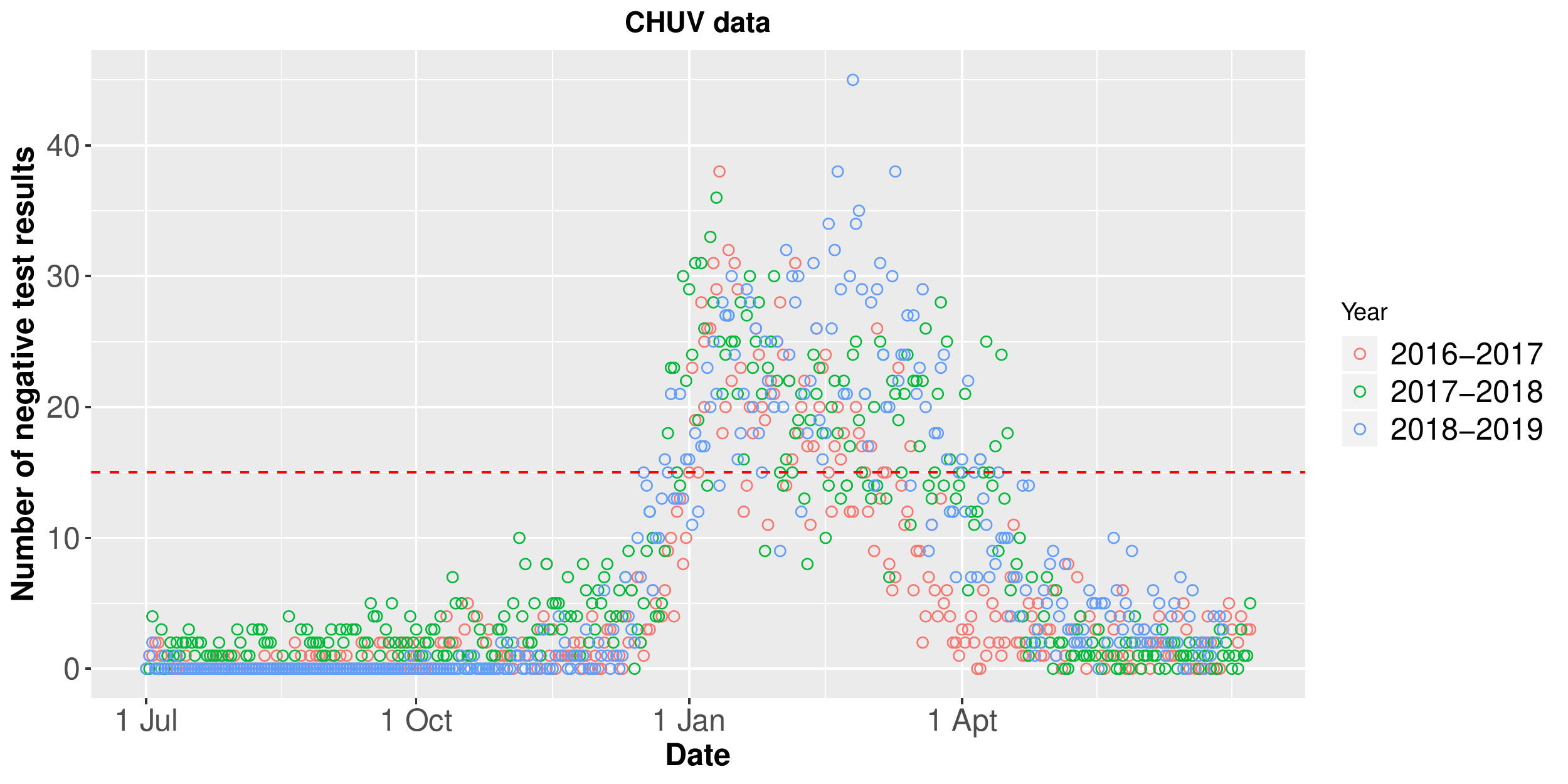}
        \caption{Daily umber of negatively flu tested patients from 01/07/2016 to 21/06/2019. The red line shows the threshold defining exceedances.}
            \label{fig:raw-negative}
\end{figure}

Figure \ref{fig:raw-positive} shows the number of positive cases for the three years considered in this study. Recording a case of flu is per se an extreme event in the sense there are generally no flu cases registered on a "normal" day. This justifies the use of exceedances of positive cases over the threshold of 1, and then to model these 290 exceedances using extreme value theory. Although flu positive cases are registered roughly between November and May, the epidemic appears at a different time each year, with patterns that differ across the years. Similar considerations apply to the negative cases shown in Figure \ref{fig:raw-negative}, where the chosen threshold is 15, which gives us 230 exceedances. The choice of threshold corresponds to a certain level of inefficiency of the flu testing process (each test is expensive and costs 180 Swiss francs). It is not economically desirable to have too many negative cases. Consider, for example, the current case of SARS-CoV-2. Some countries do not have enough testing kits and those that are available tend to be used to detect/confirm positive cases. The different patterns observed across years suggest that the calendar day variable is not a good predictor for use within the framework of hospital management of flu (positive or negative) cases. This may be in part due to meteorological variation across years and perhaps also to climate change. As for the latter, we do not have enough data to test for a long term effect. Regarding the meteorological aspect, we propose to build a model for non-identically distributed discrete extremes where covariate effects can be accounted for. The discrete generalized Pareto distribution (D-GPD) provides a theoretically justified law for discrete extremes \citep{hitz2017}, whereas, in the same spirit of generalized additive models for location, scale and shape \citep{RigbyStasinopoulos2005}, distributional parameters are made dependent on meteorological effects. The estimation approach needs to account for outlying observations as elaborated further in the next paragraph.  

\begin{figure}[ht!]
    \centering
    \includegraphics[width=0.8\linewidth]{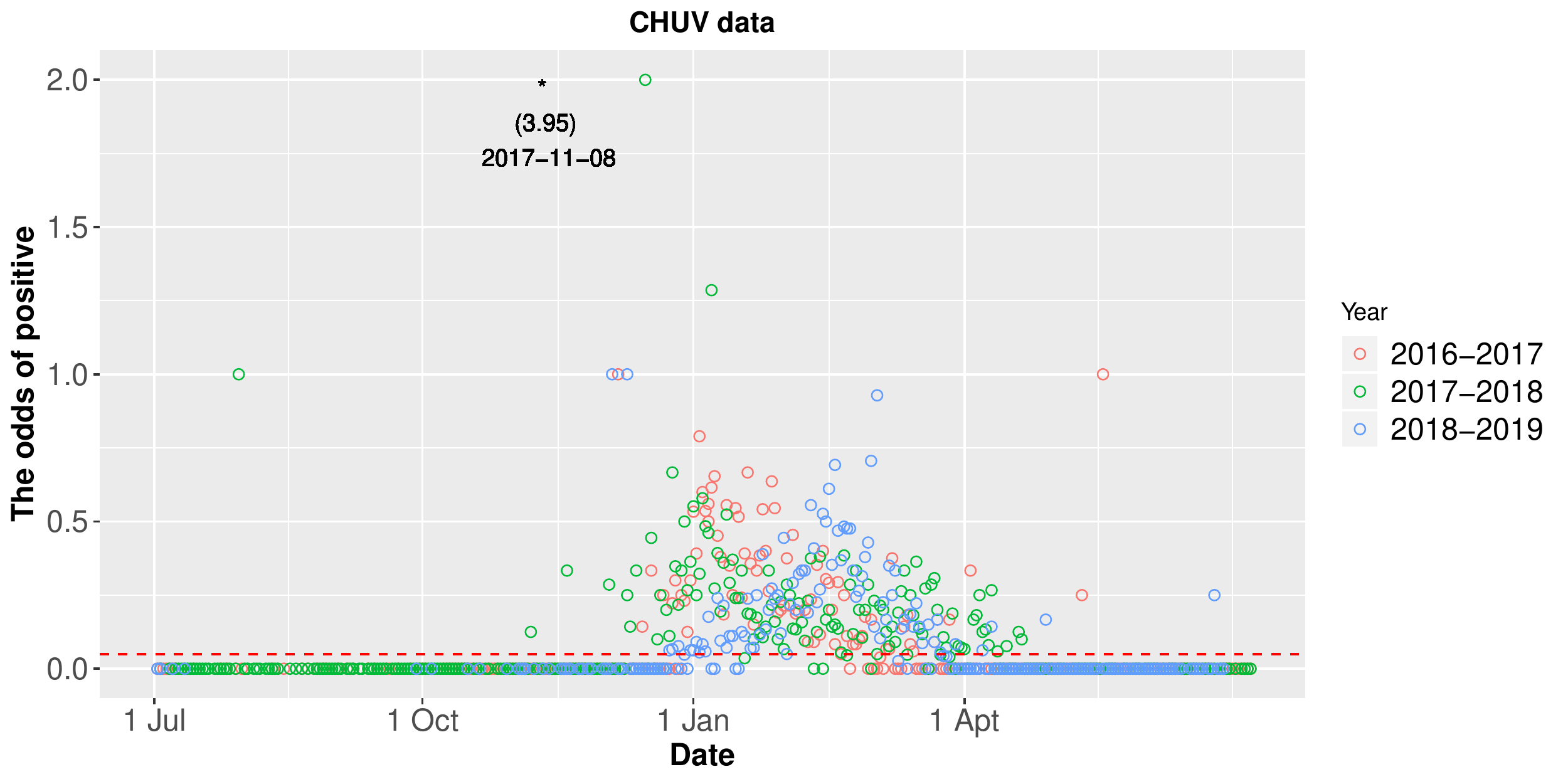}
         \caption{Daily odds of positive tested cases from 01/07/2016 to 21/06/2019. The red line shows the threshold defining exceedances.}
              \label{fig:raw-odds}
\end{figure}

From the point of view of the management of hospital congestion due to an epidemic, a measure of interest is the odds of positive cases as shown in Figure \ref{fig:raw-odds} (recall that this is the ratio between the number of positive and negative cases). This value is continuous and a high number reflects a state of epidemic which in turn means that the hospital has to prepare for congestion. We define these odds to be high if they are above a threshold of 0.05, a value that is justified by a graphical technique explained in Section \ref{evt}. The number of threshold exceedances in this case is 280. As shown in Figure \ref{fig:raw-odds}, some isolated extreme values appear outside the periods of flu epidemic. These points are found either at the very beginning of the flu period (a bell of alarm for hospitals since there are far more positive than negative cases) or at the end of the epidemic in early spring (corresponding to residual cases). Based on classical extreme value theory (EVT) the Generalized Pareto distribution (GPD) is the appropriate distribution to model such extremes and we build a model based on the GPD where the distributional parameters depend on meteorological factors. It is important to stress that testing patients for flu in hospital is a process that requires human intervention. As such, the recording process on certain days (e.g., 31st of December) will be different as compared to that of other days. This contaminates the underlying distribution of the data by creating outliers which have to be dealt with. To this end, we adapt the methodology introduced by \cite{aeberhard2019} to the specific context of this paper. To the best of our knowledge no previous work has attempted to build extreme value models based on GPD and D-GPD, where the distributional parameters are allowed to be specified as functions of covariates and the presence of outlying observations is accounted for by using a theoretically founded robust estimation approach. The newly introduced models are available through the \texttt{GJRM R} package \citep{Rpackage_GJRM} which greatly simplifies the implementation of our approach, making it as simple as a canned procedure. 


When we apply our proposed approach to the hospital data, we find that our three responses of interest depend on meteorological conditions. Specifically, our results suggest that the risk of congestion (extreme positive cases) increases when temperatures go down, and in periods of no sun and no rain. The risk of testing inefficiently (extreme negative cases) significantly increases in periods of no sun and no rain. The risk of outbreak (extreme odds of positive cases) increases in cold periods. We quantify these results in Section \ref{sct:models}.

In Section \ref{sct:method}, we describe the proposed robust regression methodology for extremes. Section \ref{sct:models} reports the results of the empirical analysis for the flu hospital data while Section \ref{sct:simulation} presents the findings from a simulation study. We conclude in Section \ref{sct:conclusion}.

%% file: 02-Method.ch
Extreme value theory is the field of statistics dedicated to the study of events with low occurrence frequencies and large amplitudes. Such events are rare in relation to the bulk of a population, which makes them hard to model and difficult to predict. Section \ref{evt} discusses the concepts of peaks-over-threshold and Charge-at-Risk-estimation (CaRe), and describes the approach used to derive the extreme distribution used to analyze the discrete outcomes of this paper. Section \ref{sct:gamlss} discusses the robust approach used to estimate GPD and D-GPD models whose distributional parameters are allowed to depend on covariate effects.

\subsection{GPD, D-GPD and CaRe}
\label{evt}

One of the classical theories of extremes for a common continuous random variable is based on the probabilistic limit result for exceedances of high thresholds. The so-called peaks-over-threshold (POT) consists of a limiting result for the times and sizes of exceedances over some high level. Letting $(Y_{i})_{i \geq 1}$ be a sequence of independent and identically distributed (iid) random variables in $[0, y_F)$, with common continuous distribution $F$, we concentrate on the sizes of exceedances over some high threshold. In other words, the focus is on the right tail of distribution $F$. The result by \cite{balkema1974} allows for the approximation of the conditional distribution of the exceedances above a high threshold $u$. If there exist normalizing sequences $\{a_n>0\}$ and $\{b_n\}$ such that for $u \to y_F$,
\begin{eqnarray}
a_n^{-1}(Y-u) & \vert  Y\geq u \overset{d}{\to} Z
\label{MDAcondition}
\end{eqnarray}
for some $Z$ following a non-degenerate probability distribution on $[0, \infty)$, where $\overset{d}{\to}$ denotes weak convergence, then it is possible to model the limiting distribution of the exceedances $Y-u\vert Y>u$ with a Generalized Pareto distribution (GPD), that is,
\begin{equation}
\Pr\left( Y> u+y \mid Y >u \right) \underset{u\to y_F}{\longrightarrow} 
G_{(\sigma,\xi)}(y) = \left\{
 \begin{array}{ll}
 \left( 1+\xi y/\sigma\right) _{+}^{-1/\xi}, & \xi \neq0, \\
 \exp(-y/\sigma), & \xi =0,
 \end{array}
\right. \label{gpd_asymptotic}
\end{equation}
where $\xi\in \R$ is the shape parameter and $\sigma>0$ the scale parameter. The case $\xi=0$ is interpreted as the limit. Parameter $\xi$ provides information about the heaviness of the tail of the underlying distribution $F$. More formally, condition (\ref{MDAcondition}) means that $Y$ belongs to the maximum domain of attraction of an extreme value distribution with shape parameter $\xi$ and we write $Y\in \mbox{MDA}(\xi)$. Some examples are the Pareto distribution,
$$F(y)=1-\left(\frac{\kappa}{\kappa+y}\right)^{\alpha}\,,
\quad\alpha,\kappa>0,\quad x\geq0, $$
for which we take $a_n=\kappa
n^{1/\alpha}/\alpha$, $b_n=\kappa n^{1/\alpha}-\kappa$, and is in MDA($1/\alpha$) (Fr\'echet case). The exponential distribution,
\mbox{$F(y)=1-e^{-\lambda y}$}, \mbox{$\lambda>0$}, \mbox{$y\geq0$}, for which we take $a_n=1/\lambda$, $b_n=(\log n)/\lambda$, and is in MDA($0$) (Gumbel-case).

Essentially all commonly encountered continuous
distributions are in the maximum domain of attraction of an extreme value distribution. If the tail of $F$ decays like a power function then $F$ is in MDA($\xi>0$). Distributions such as Burr, log-gamma, Cauchy, Pareto and Student-t as well as various mixture models are heavy-tailed. The Gumbel class characterized by $\xi=0$ contains light-tailed distributions such as the Gaussian, log-normal, exponential and gamma whose tail decay roughly exponentially. The so-called Weibull class, defined by $\xi<0$, contains distributions that are bounded above (e.g., uniform and Beta distributions). In other words, for a wide class of distributions, the distribution of the excesses over a high threshold can be approximated by the GPD. This result suggests that if we choose $u$ high enough then we can assume that result (\ref{gpd_asymptotic}) holds for some parameters $\xi$ and $\sigma$. In practice, such parameters are estimated by fitting a GPD to the excess amounts over the threshold $u$, relying on the standard properties of maximum likelihood estimators for $\xi > -0.5$.

The choice of $u$ is important; a very low threshold would lead to high bias whereas a very high threshold to high variance. For practical purposes, a suitable threshold $u$ is chosen using data-analytic tools such as the mean residual life plot. This is based on the fact that if excess losses over threshold $u$ can be characterized by a GPD with parameters $\xi <1$ and $\sigma$, then it is easy to show that for any higher threshold $v \geq u$
\begin{eqnarray}
E(Y-v \mid Y>v) & = & \frac{\sigma+\xi (v-u)}{1-\xi},
\label{mrplot}
\end{eqnarray}
so that the mean excess function is linear in $v$ above $u$. In empirical applications, the mean residual life plot (which shows the empirical mean excess against increasing threshold values) is a useful tool to choose the threshold and also to determine the adequacy of the GPD model as an approximation of the excess distribution. For a linear mean excess function characterizing the GPD class, the best threshold candidate is any value for which the mean excess looks linear. Figure \ref{fig:mrl.plot} shows mean residual life plots for the odds of positive cases without removing outliers (top panel) and after removing 1\% of the largest values (bottom panel). The graphs support the presence of a contamination effect due to outliers, which may indeed make the choice of the threshold difficult. This is standard for mixed distributions.

Taking the point of view of the hospital's risk management, an important quantity is the quantile of $Y$ given that $Y>u$ because it quantifies the information about the charge load the hospital has to be ready for under different scenarios of congestion. In practice, for a fixed threshold $u$, given that $Y>u$, and a horizon of $h$ days, the $p$-quantity, called Charge-at-Risk-estimation (CaRe), is the value of $Y$ that might be exceeded one time in $h$ days. This is defined as
 \begin{eqnarray}
 \mbox{CaRe}(p)_{\mbox{GPD}}  & = & u+ \frac{\sigma}{\xi}\left\{(1-p)^{(-\xi)}-1\right\},
 \label{eq:CaRe_GPD}
 \end{eqnarray}
 where $h=1/(1-p)$. As an example, during a flu period ($Y>u$), 97\%-CaRe roughly corresponds to the value of $Y$ that might be exceeded once in a month. In a financial context, a related value is the co-called Value-at-Risk imposed by the Basel committee and used, for instance, to measure the risk of loss on a specific portfolio of financial assets.  

For a discrete random variable $R$, using the GPD to approximate the distribution's tail behaviour can be inappropriate. As pointed out by \cite{hitz2017}, many common distributions such as the Poisson, geometric and negative binomial are not in any maximum domain of interest. \cite{hitz2017} proposed two methods for modelling the tails of discrete observations from distributions with infinite support. In this paper, we briefly recall one of them. Defining the discrete maximum domain of attraction as D-MDA, we write $R\in$D-MDA($\xi$) with $\xi\geq 0$ if there exists a continuous random variable $Y$ such that $P(R \geq r) = P(Y \geq r)$ for $r = 0,1,2, \ldots$. Then, for large integers $u$, we have
 $P(R -u = r \vert R \geq u) = P(Y -u \geq r \vert Y \geq u) - P(Y -u \geq r + 1 \vert Y \geq u)$, which, from (\ref{gpd_asymptotic}), tends to a discrete generalized Pareto distribution (D-GPD) defined by
 \begin{eqnarray}
 DG_{(\sigma,\xi)}(r) & =& \bar{G}_{(\sigma,\xi)}(r) - \bar{G}_{(\sigma, \xi)}(r+1),
 \label{dgpd}
 \end{eqnarray}
for $r = 0,1,2, \ldots$, where $\bar{G}$ denotes the survival function of $G$ (see \cite{hitz2017} and references therein).

	
	In risk management, as mentioned ealier, the quantile is a quantity of interest. In what follows, we derive such an expression for the D-GPD. Let $p \in (0,1)$ be the probability for which we seek a quantile, we then solve
  	
  	\begin{align*}
  	p & = \sum_{r=0}^{q} \left(1+\frac{\xi r}{\sigma}\right)^{(-1/\xi)} - \sum_{r=0}^{q} \left(1+\frac{\xi(1+r)}{\sigma}\right)^{(-1/\xi)}\\
  	&= 1 + \sum_{r=1}^{q} \left(1+\frac{\xi r}{\sigma}\right)^{(-1/\xi)} -\sum_{z=1}^{1+q}\left(1+\frac{\xi z}{\sigma}\right)^{(-1/\xi)}= 1-\left(1+\frac{\xi(1+q)}{\sigma}\right)^{(-1/\xi)}.
  	\end{align*}
 
 As in (\ref{eq:CaRe_GPD}) for the continuous context, the $p$\%-CaRe for the discrete case is
 \begin{eqnarray}
 \mbox{CaRe}(p)_{\mbox{D-GPD}}  & = & \left\lceil u+ \frac{\sigma}{\xi}\left\{(1-p)^{(-\xi)}-1\right\}\right\rceil-1,
 \label{eq:CaRe_DGPD}
 \end{eqnarray}
 where $\lceil ~~\rceil$ denotes the ceiling function (the smallest integer greater than or equal to).
 
The next section provides details on the robust estimation approach employed to fit extreme value models based on the GPD and D-GPD, where $\sigma$ and $\xi$ can be specified as functions of covariate effects.



\subsection{Covariate Effects and Parameter Estimation}\label{sct:gamlss}

In the context of POT models for continuous variables through GPD excess size approximations, employing techniques that allow for flexible forms of dependence on covariates are very attractive in empirical applications \citep{Davison}. To this end, \cite{chavez2005} employed the framework of generalized additive models \citep{hastie1990, Wood17} to estimate flexibly the shape and scale parameters of an orthogonal reparametrization of the GPD. \cite{Yee2007} proposed instead the use of vector generalized additive models. These modelling strategies are philosophically consistent with generalized additive models for location, scale and shape \citep{RigbyStasinopoulos2005}, where, for any continuous or discrete distribution $F_{\boldsymbol{\theta}}$, with $\boldsymbol{\theta}$ being a $d$-dimensional parameter vector with virtually any $d>0$, all distributional parameters are allowed to depend on covariate effects. This type of modelling has received a great deal of interest since its introduction and some researchers also refer to it as distributional or multi-parameter regression. The classical and perhaps most commonly known software implementation of such models is the \texttt{gamlss R} package \citep{RigbyStasinopoulos2005}. Another implementation is available via the \texttt{gamlss()} function from the \texttt{GJRM R} package \citep{Rpackage_GJRM} which has been extended to incorporate the models developed in this paper. 




For $i=1, \ldots,n$, where $n$ denotes the sample size, let $Y_i$ be independently sampled from $F_{\boldsymbol{\theta_i}}$ (with density or probability function $f_{\boldsymbol{\theta_i}}$), where $\boldsymbol{\theta_i} = (\theta_{i1}, \ldots, \theta_{id})$ and $\boldsymbol{x}_i$ is a vector of covariates of dimension $p$ (which can include binary, categorical, and continuous variables, for instance). The distributional assumption of $Y_i$ is understood to be conditional on all covariates. This is achieved by assuming for each parameter $\theta_{ij}$, for $j=1,\ldots,d$, that
\begin{equation}\label{eq:gamlss}
g_j\left(\theta_{ij} \right)    = \beta_{j0} + f_{j 1}(\boldsymbol{x}_{i j 1}) + \ldots +f_{j k}(\boldsymbol{x}_{i j k}) +  \ldots  + f_{j K_j}(\boldsymbol{x}_{i j K_j}), 
\end{equation}
where the $g_j$ are one-to-one transformations or link functions (ensuring that the parameters range restrictions are met), $\beta_{j0}\in \mathbb{R}$ are overall intercepts, $\boldsymbol{x}_{i j k}$ denotes the $k$th sub-vector of covariates pertaining to term $j$ and observation $i$, and the $K_j$ functions $f_{j k}(\cdot)$ represent generic covariate effects (which can be of any pre-specified parametric form such as linear or quadratic, or can be non-parametric). Each of these functions are approximated by a linear combination of $J_{k j}$ basis functions $b_{k j l}(\boldsymbol{x}_{i k j})$ and regression coefficients $\beta_{k j l }\in\mathbb{R}$, that is, $f_{j k}(\boldsymbol{x}_{i j k}) \approx \sum_{l=1}^{J_{k j}} \beta_{d k j} b_{k j l}(\boldsymbol{x}_{i k j})$. This (regression spline) approach allows for a vast variety of covariate effects. We refer the reader to \cite{Wood17} for all the options available and that are supported by our implementation. Note that, in our case study, linear specifications were deemed to be sufficient to model the variation in the response variables of interest.  

For GPD and D-GDP, we have that $\boldsymbol{\theta_i}  =  (\theta_{i1}, \theta_{i2}) = (\xi_i, \sigma_i)$, hence $d=2$. The choices of one-to-one transformations have to guarantee that the parameters lie in their admissible definition spaces. For GPD, with probability function defined as in Equation~\eqref{gpd_asymptotic}, we have  
$$
g_1(\xi_i) =  \log\left( \xi_i + 0.5 \right) \mbox{ and } g_2( \sigma_i) = \log \left( \sigma_i \right).
$$
For D-GDP, with probability function given by Equation~(\ref{dgpd}), we employ 
$$
g_1(\xi_i) = \sqrt{\xi_i} \mbox{ and } g_2( \sigma_i) = \log \left( \sigma_i \right).
$$
The above choices ensure that $\sigma_i$ is positive for both distributions, that $\xi_i > -0.5$ for GPD (otherwise parameter estimation is non-regular in the sense that the score statistic is not asymptotically normal as argued by \cite{Davison}), and that $\xi_i > 0$ for D-GDP as required \citep{hitz2017}.

Let $\boldsymbol{\delta}$ be the vector of the model's parameters to be estimated. This includes the coefficients associated with (\ref{eq:gamlss}). Model fitting is performed by maximizing the log-likelihood function $\ell(\boldsymbol{\delta}) = \sum_{i=1}^n \ell(\boldsymbol{\delta})_i  = \sum_{i=1}^n \log f\left(y_{i}| \boldsymbol{\theta}\right)$. Note that, although not required for our case study, our implementation supports the presence of non-parametric components. In this case, the objective function would be augmented by a penalty term defined as $1/2 \boldsymbol{\delta}^{'} \mathbf{S}\boldsymbol{\delta}$, where $\mathbf{S}$ is a matrix that depends on the choice of basis functions for the non-parametric terms, and on a set of smoothing parameters that controls the trade-off between fit and smoothness.

If outlying observations occur in the data, classical model fitting will suffer from a lack of robustness, which will adversely affect parameter estimates. To deal with this, we adopt the methodology of \cite{aeberhard2019} which essentially consists of reducing the likelihood contributions of low log-likelihood values while leaving large log-likelihood evaluations essentially unchanged. This is achieved through a function $\rho_c$ applied to the log-likelihood components, so that the objective function becomes $\tilde{\ell}(\boldsymbol{\delta}) = \sum_{i=1}^n \rho_c\big(\ell(\boldsymbol{\delta})_i\big) - b_{\rho}(\boldsymbol{\delta})$, where 
$$
b_{\rho}(\boldsymbol{\delta}) = \sum_{i=1}^n b_{\rho}(\boldsymbol{\delta})_i = \sum_{i=1}^n  \int \rho_c^\star\big(\log f(y|\boldsymbol{\delta})\big) \, \mathrm{d} y
$$ 
is a correction factor ensuring Fisher consistency, and $\rho_c^\star$ is directly derived from the specified $\rho_c$ through $\rho_c^\star(z) = \int_{-\infty}^z \exp(s) \rho_c'(s) \, \mathrm{d} s$ with $\rho_c'(s) = \partial \rho_c(s) / \partial s $. 

The tuning constant $c>0$ in $\rho_c$ regulates the trade-off between loss of estimation efficiency (should the data exactly come from the assumed model) and the magnitude of the maximum estimation bias (should the data not come from the postulated model). For any given $c$, $\rho_c$ is assumed to be convex, monotonically increasing and twice continuously differentiable over $\mathbb{R}$, and to have bounded first derivative $\rho_c'$ within $[0,1]$. The latter can be interpreted as a multiplicative robustness weight, as one would add when weighting the estimating equations in robust $M$-estimation. An advantage of the approach is that it leads to a natural definition of robust information criteria. 

Regarding the choice of $\rho_c$, \cite{aeberhard2019} recommend using the log-logistic function first proposed by \cite{EguchiKano2001}: $\rho_c(z) = \log\frac{1+\exp(z+c)}{1+\exp(c)}$, for $c>0$, with corresponding $\rho_c^\star(z) = \exp(z) - \exp(c)\log\big(1+\exp(z+c)\big)$ and first derivative $\rho_c'(z) = \exp(z+c)/\big(1+\exp(z+c)\big)$. It holds that $\lim_{c\rightarrow\infty}\rho_c(z) = z$ so that an increasingly large $c$ value leads to the (non-robust) original $\ell(\boldsymbol{\delta})$. The value of $c$ is tuned via a simulation based procedure that controls how the robustness weights at the score level (represented by $\rho_c^{\prime}$) behave under data generated from the assumed model. The user can decide the level of down-weighting to be achieved with respect to maximum likelihood (e.g., $95\%$).

\cite{aeberhard2019} established the Fisher consistency of $\hat{\boldsymbol{\delta}}$ as well as its asymptotic Gaussian distribution and asymptotic variance-covariance matrix which can be used to construct confidence intervals. The authors discussed a Bayesian inferential result as well. This is advantageous because such a result does not rely on asymptotic considerations, and intervals for non-linear functions of the model's parameters (e.g., CaRe) can be reliably and efficiently obtained via posterior simulation. The adopted estimation framework allows for the elegant construction of robust information criteria such as the robust AIC, that is, 
\begin{equation}
\label{eq:raic}
\text{RAIC}(\boldsymbol{\lambda}) = -2 \tilde{\ell}(\boldsymbol{\delta}) + 2\texttt{edf},
\end{equation}
where \texttt{edf} denotes the effective degrees of freedom which is given by the trace of a matrix that depends on components of the asymptotic variance-covariance matrix of $\hat{\boldsymbol{\delta}}$ \citep{aeberhard2019}.

In order to estimate the model's coefficients, we have extended the efficient and
stable trust region algorithm proposed by \cite{MR2019} to our context. One of the many advantages of such an algorithm is that it does not require the orthogonality of the distributional parameters (in this case, $\xi$ and $\sigma$). The implementation of the trust region approach required the analytical score and Hessian of the model's log-likelihood which were derived and are reported in Supplementary Section \ref{sct:appex}. 

Although robust models for extremes have been developed in the literature \citep[see, e.g.,][]{Dupuis.1998, Dupuis.2006, DellAquila}, we would like to stress that, to the best of our knowledge, there are no alternative robust regression models for extreme distributions, nor respective software implementations, of the type discussed in this paper. While the construction and estimation of the proposed model rely on the infrastructure and modelling framework of \texttt{GJRM}, extending the software to accommodate the developments needed to address the challenges of our case study required a great deal of work. Supplementary Section \ref{sct:appexSoft} provides details on the usage of function \texttt{gamlss()} from the \texttt{GJRM R} package.

%% file: 03-Models.ch
\label{sct:models}
Flu is contagious and it can spread by airborne respiratory droplets, saliva or skin-to-skin contact and by touching a contaminated surface. In Switzerland, a sentinel surveillance system and a mandatory reporting system are used to register flu data. Flu monitoring in hospitalized patients has also been in a testing phase since 2018. From a health care managerial point of view, the number of negative cases among tested patients showing flu-like symptoms is as important as the number of positive cases. An extreme number of positive cases may indicate an epidemic of such a viral infection. An extreme number of negatives can instead indicate management inefficiency and hence excessive financial costs as well as congestion in the hospital units that are involved in diagnostics and pre-treatment. The contagion aspect of hospitalized patients also needs to be taken into account in the bed organization of the hospital. 

As shown in Figures \ref{fig:raw-positive}, \ref{fig:raw-negative} and \ref{fig:raw-odds}, the annual calendar variable does not seem to provide important insights into managing and/or preventing congestion due to a flu epidemic. Based on the literature on flu transmission, we use meteorological variables for modelling the extremes of: positive flu cases, negative cases (visiting the hospital for a flu check) and the odds of positives. The meteorological variables are represented by $\operatorname{L}^3 X_{t}=X_{t-3}$, where $\operatorname{L}$ is the usual lag operator and $X_{t}$ can be each of the variables in Table \ref{tab:cov}. For each meteorological factor, we consider the respective $\operatorname{L}^3 X_{t}$ value, because, for flu, the incubation time is usually between 24 and 48 hours, and sometimes 72 hours. When discussing the models used for the analyses, extension \texttt{L3} at the end of each covariate's name refers to $\operatorname{L}^3 X_{t}$ values.

\begin{table}[H]
    \centering
    \begin{tabular}{llc}
    \hline
       Name   & Definition & Unit \\\hline
       {\bf Mintemp}  & daily minimum temperature at 2m above ground level & $^o C$ \\
       {\bf  Radiation} & daily mean  radiation&   $W/m^2$\\
       {\bf Humidity} & daily mean relative air humidity at 2 m from the ground  & \% \\
       {\bf Wind} & daily maximum wind (integration 1 s) & $m/s$\\
       {\bf  Precipitation} & daily sum of precipitation &  $mm$\\
       {\bf Pressure} &  daily mean atmospheric pressure with (QNH)& $hPa$\\\hline 
    \end{tabular}
    \caption{Potential meteorological covariates.}
    \label{tab:cov}
\end{table}

\begin{figure} [H]
    \centering
    \includegraphics[width=0.8\linewidth]{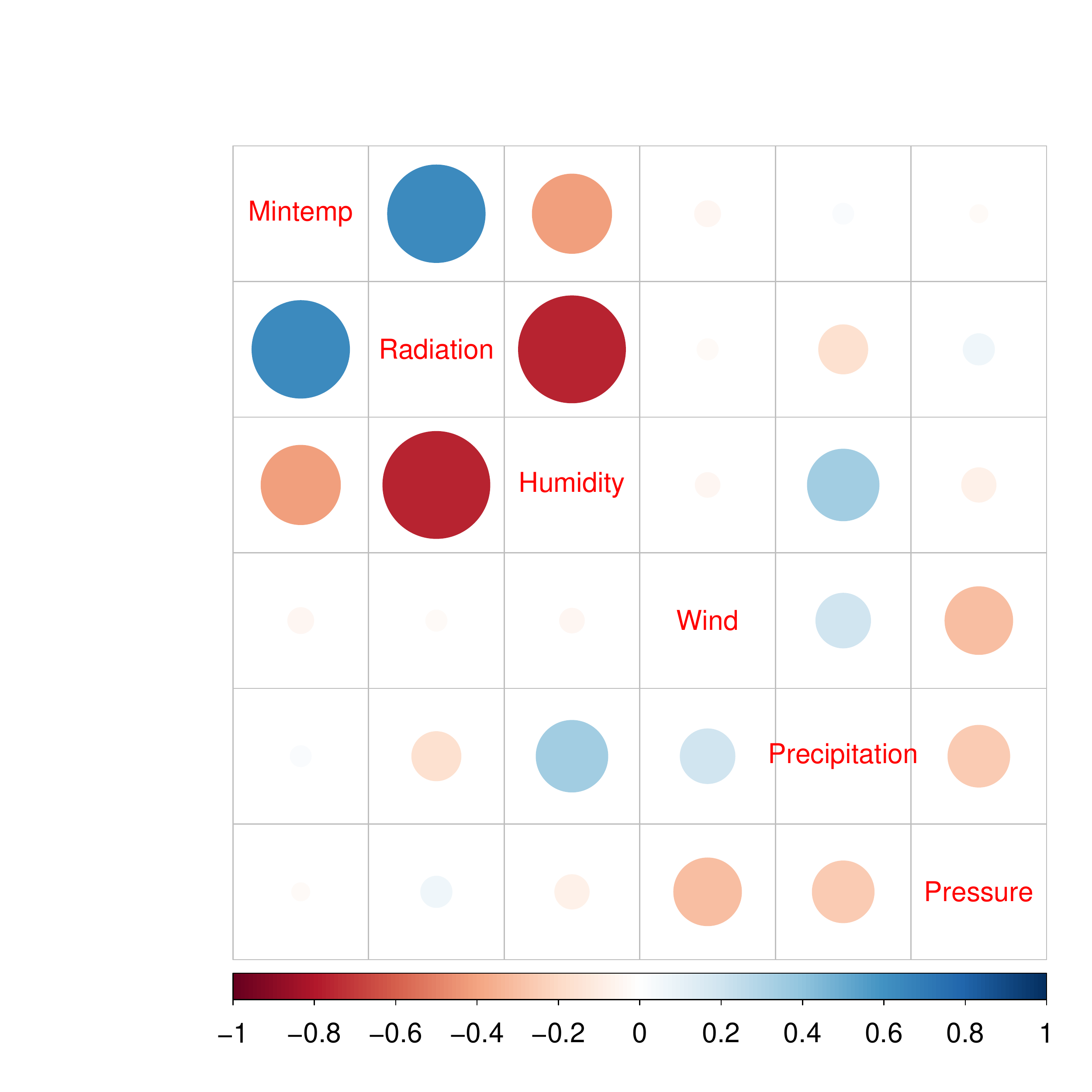}
        \caption{Correlation between meteorological covariates.}

    \label{fig:corrplot}
\end{figure}

As for the flu data, the meteorological variables were measured in Lausanne (Switzerland) from July 1, 2016 to June 21, 2019 and are available at https://gate.meteoswiss.ch/idaweb/. Figure \ref{fig:corrplot} shows a correlation plot between the meteorological variables, highlighting, as expected, that humidity and radiation are highly and negatively correlated, whereas radiation and minimum temperature are positively correlated. Recall that the aim is to quantify the effect of meteorological factors on the extremes of positive, negative, and the odds of positive cases. For these responses, we specify three different models (estimated using the approach described in Section \ref{sct:gamlss}) based on the D-GPD for the two discrete responses and on the GPD for the odds of positive extremes. The models robustness tuning constants $c$ were set to 6.1, 6.7 and 2.6, respectively, to achieve a level of down-weighting of $95\%$.

\begin{figure} [H]
    \centering
    \includegraphics[width=0.8\linewidth]{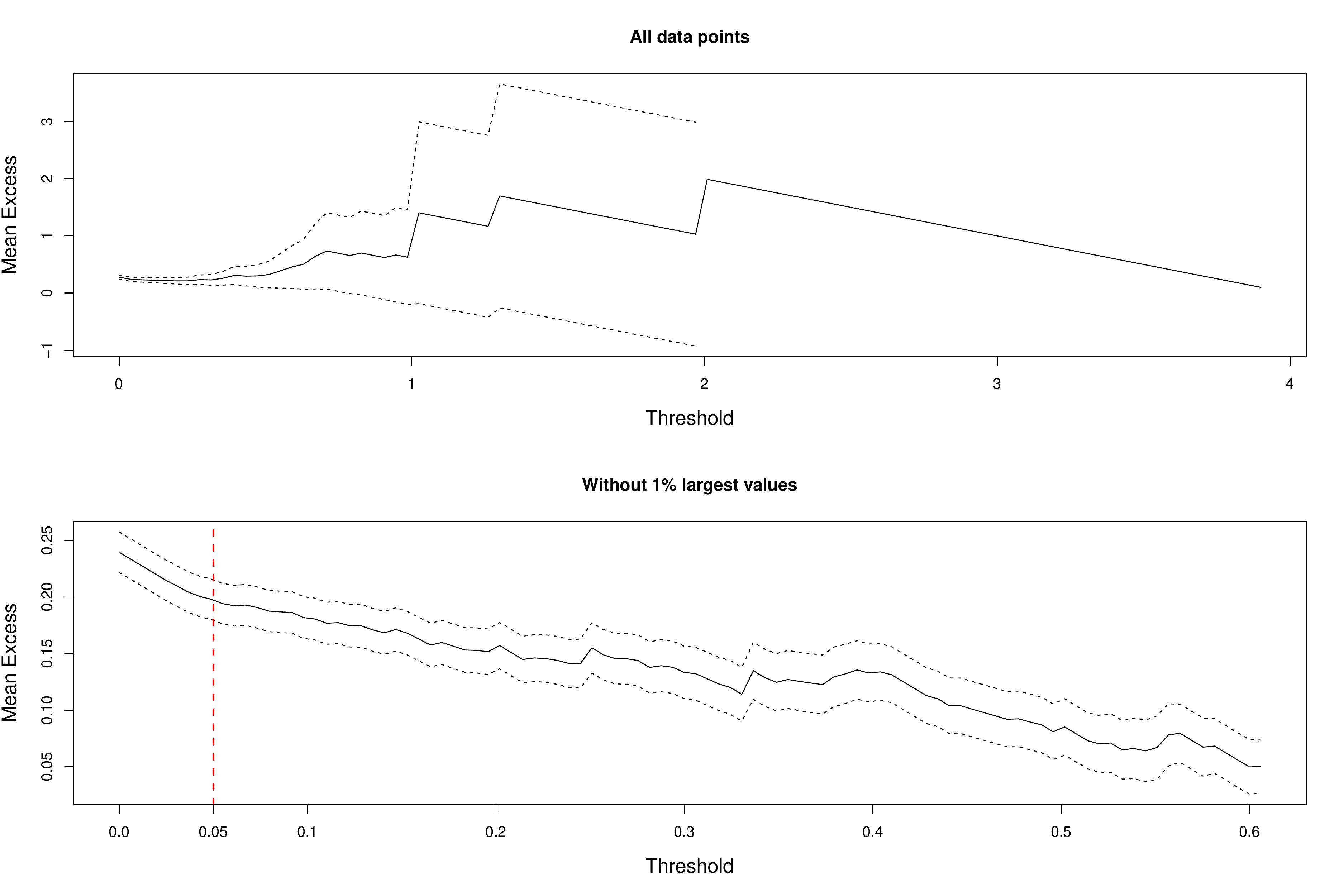}
    \caption{Mean residual life plots for the odds of positive cases.}
     \label{fig:mrl.plot}
\end{figure}

Since the odds of positive cases is a continuous variable, we employed the mean residual life plot, based on (\ref{mrplot}), to choose the threshold defining the tail of the underlying distribution. Figure \ref{fig:mrl.plot} shows such a plot when using all the odds data (upper panel) and without 1\% of the largest values (lower panel). Looking at the upper panel, it is difficult to decide the threshold value: the mean excess curve exhibits very little linear stability and the confidence bounds become increasingly wide. In this case we choose 0.05, the value for which the mean excess tends to increase in a roughly linear manner. The lower panel, which excludes potential outliers, confirms the adequacy of the chosen threshold value. It is interesting to note that, in agreement with the analysis of \cite{neslehova2006} for mean excess plots in the presence of contaminated data, the results shown in Figure \ref{fig:mrl.plot} support the presence of contamination due to outlying observations. More precisely, the different nature of the largest values makes the mean excess curve of the upper panel difficult to interpret.

Forward variable selection (based on the RAIC defined in Equation \eqref{eq:raic}) was performed for the three models. We checked for linear and non-linear covariate effects; as mentioned in Section \ref{sct:gamlss}, linear specifications were found to be adequate for the modelling purposes of our dataset.  

For all responses, a constant model was selected for the shape parameter $\xi$. This is not surprising, first, because this parameter is difficult to estimate \citep{Hosking87}, and, second, because it concerns the tail behavior of the underlying process, a characteristic that is not likely to vary with meteorological factors. The estimated shape parameters related to the two discrete variables (positive and negative) are close to zero, meaning an underlying light tail that in fact many common discrete distributions, including
geometric, Poisson and negative binomial distributions have \citep{hitz2017}. The estimated shape parameter for the odds of positive has a negative sign, which makes sense as one can expect this odd to be bounded above. As for the scale parameter $\sigma$, we found significant and linear effects for some meteorological variables. Below we present the results for each response variable.

\underline{Number of positive cases}\\
In this case, the final fitted D-GPD model is based on the following equations and estimates: 
$$ 
\sqrt{\hat{\xi}}=\hat{\alpha}_{0},
$$
with $\hat{\alpha}_{0} = -1.974$e-06 (and standard error of $3.719$e-01) and 
$$ 
\log(\hat{\sigma}) = \hat{\beta}_{0} + \hat{\beta}_{1}\texttt{MintempL3} + \hat{\beta}_{2}\texttt{RadiationL3} + \hat{\beta}_{3} \texttt{PrecipitationL3}.
$$
The estimated coefficients are reported in Table~\ref{tab:positive-models} and their effects graphically shown in Figure \ref{fig:positive-eq2} in Supplementary Section~\ref{sct:appexSupp}.
\begin{table}[H]
\centering
\caption{Estimated coefficients for the final D-GPD model fitted to positive cases.}
\begin{tabular}{lccccc}
\hline
  $\log(\hat{\sigma}) $ & Estimate & Std. Error & Z- value & P-value & Signif \\\hline
Intercept       & $2.037$    & $0.138$      & $14.79$    & $0.000$   & *** \\\hline
\texttt{MintempL3} & -0.054   & 0.021      & -2.54    & 0.011   & *   \\\hline
\texttt{RadiationL3}       & -0.005   & 0.001      & -3.62    & 0.000   & ***    \\\hline
\texttt{PrecipitationL3}   & -0.009   & 0.004      & -2.11    & 0.035   & *     \\\hline
\multicolumn{6}{l}{Signif. codes:  $0 < \mbox{p-value}<1e^{-3}: ***$, $\mbox{p-value}<0.01: **$, $\mbox{p-value}< 0.05: *$}
\end{tabular}
\label{tab:positive-models}
\end{table}
The resulting estimated value for $\xi$ is very close to zero, implying a probability mass function that may belong to the discrete maximum domain of attraction with $\xi=0$ \citep{hitz2017}. The equation for the scale parameter $\sigma$ explains both the variable's variance and mean. For the latter, for GPD and therefore also for D-GPD, recall that the mean excess depends on the scale parameter as shown in Equation~(\ref{mrplot}). A broad interpretation of the results is that the warmer and nicer the weather, the lower the number of extreme positive cases, variability and mean. Interestingly, radiation seems to better explain the response than humidity does, which is the factor ommonly used to explain flu spread \citep{lowen2014}. As highlighted by Figure~\ref{fig:corrplot}, radiation and precipitation provide complementary proxies of humidity.  

\underline{Number of negative test results}\\
The final fitted D-GPD model is based on the following equations and estimates:
$$ 
\sqrt{\hat{\xi}}= \hat{\alpha}_{0},
$$
with $\hat{\alpha}_{0} =4.505$e-05 ($1.142$e-01) and  
$$ 
\log(\hat{\sigma}) = \hat{\beta}_{0} +  \hat{\beta}_{1}\texttt{RadiationL3} + \hat{\beta}_{2} \texttt{PrecipitationL3}.
$$
The estimated coefficients are reported in Table~\ref{tab:negative-models} and their effects shown in Figure \ref{fig:negative-eq2} in Supplementary Section~\ref{sct:appexSupp}.
\begin{table}[H]
\centering
\caption{Estimated coefficients for the final D-GPD model fitted to negative cases.}
\begin{tabular}{lccccc}
\hline
   $\log(\hat{\sigma})$           & Estimate & Std. Error & Z- value & P-value & Signif \\\hline
Intercept     & 2.483      & 0.151    & 16.43   & 0.000 & *** \\\hline
\texttt{RadiationL3}     & -0.003     & 0.001    & -2.15   & 0.031 & *   \\\hline
\texttt{PrecipitationL3} & -0.010     & 0.005    & -2.01   & 0.045 & * \\\hline
\multicolumn{6}{l}{Signif. codes:  $0 < \mbox{p-value}<1e^{-3}: ***$, $\mbox{p-value}<0.01: **$, $\mbox{p-value}< 0.05: *$}
\end{tabular}
\label{tab:negative-models}

\end{table}

As compared to the previous model, minimum temperature does not seem to explain the variability and mean of the number of negative extremes. This may be due to the fact that cold weather activates the virus which in turn leads to more positive cases. Both the effects of radiation and precipitation are less important than those found when modelling positive cases. This may be explained by the fact that better meteorological conditions (warmer months and sun) simply decrease the number of test cases. For positive cases, favorable meteorological conditions also decrease the probability of catching the flu during the autumn/winter time. In other words, the evidence suggests that radiation and precipitation, which commonly affect the positive and negative cases, mostly influence the total number of tests that are carried out. During warmer months, summer or when the weather is good during winter time, individuals tend not to go the hospital. The analysis suggests that minimum temperature influences only the spread of flu. 

\underline{Odds of positive cases}\\
The final fitted GPD model is based on the following equations and estimates:
$$ 
\log(\hat{\xi}+0.5)=\hat{\alpha}_{0},
$$ 
with $\hat{\alpha}_{0}= -3.293(0.671)$ and  
$$ 
\log(\hat{\sigma}) = \hat{\beta}_{0} + \hat{\beta}_{1}\texttt{MintempL3}.
$$
The estimated coefficient for \texttt{MintempL3} is reported in Table~\ref{tab:odds-models} and its effect displayed in Figure \ref{fig:odds-eq2} in Supplementary Section~\ref{sct:appexSupp}.
\begin{table}[H]
\centering
\caption{Estimated coefficients for the final GPD model fitted to the odds positive cases.}
\begin{tabular}{lccccc}
\hline
 $\log(\hat{\sigma})$    & Estimate & Std. Error & Z- value & P-value & Signif \\\hline
Intercept       & -1.272 & 0.051 & -25.03 & 0.000 & *** \\\hline
\texttt{MintempL3} & -0.098 & 0.009 & -11.17  & 0.000 & ***\\\hline
\multicolumn{6}{l}{Signif. codes:  $0 < \mbox{p-value}<1e^{-3}: ***$, $\mbox{p-value}<0.01: **$, $\mbox{p-value}< 0.05: *$}
\end{tabular}
\label{tab:odds-models}
\end{table}
In this case the resulting estimated value of $\xi$ is negative, implying that the underlying distribution is bounded above, that is, the ratio cannot reach extremely large values. 
Unsurprisingly the extremes of the odds positive do not depend on meteorological factors other than those explaining positive and negative extreme cases. However, this was not obvious because the threshold choice for the odds positive was not given by the threshold for the positive and negative cases. In this model, only minimum temperature influences the response. All in all, among the different meteorological factors observed, humidity, wind and pressure do not exhibit an effect as compared to radiation, precipitation and minimum temperature; the latter being more responsible for the epidemiological aspect of the analysis. 


\subsection{Hospital congestion} 

From a risk management point of view, a quantity of high interest is the CaRe defined in Equation~\eqref{eq:CaRe_GPD} for GPD, and in Equation\eqref{eq:CaRe_DGPD} for D-GPD. These values constitute important risk measures for the hospital. Intervals are also crucial as they provide information about the estimates' uncertainty. 

\begin{figure}[H]
    \centering
    \includegraphics[width=0.8\linewidth]{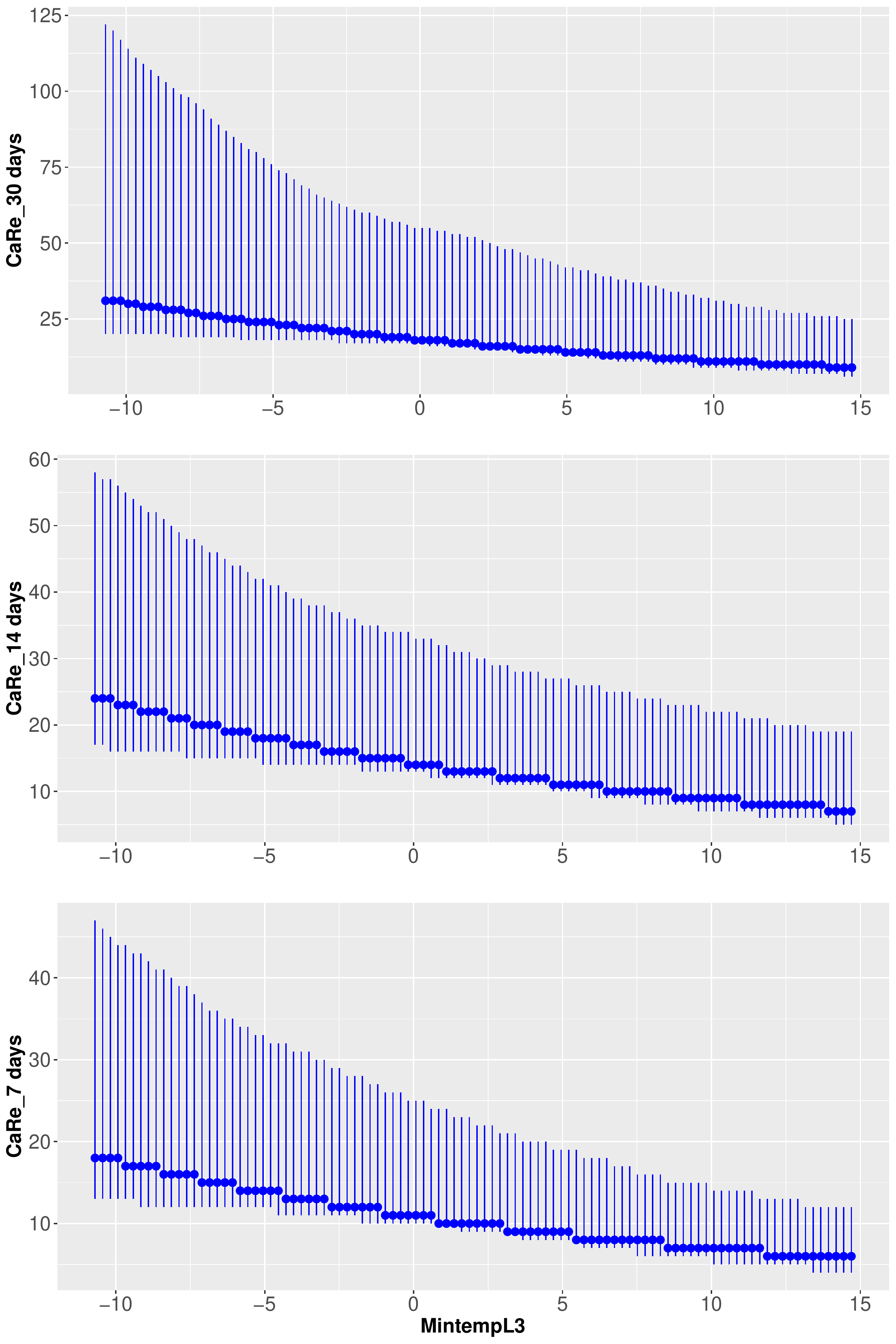}
    \caption{Positive cases: estimated CaRe for $h = 30, 14, 7$ days (panels from top to bottom, respectively) with respect to lagged minimum temperature. The bars correspond to 95\% intervals. The other covariates are fixed to their mean values.
    }
    \label{fig:Qpositive_mintemp}
\end{figure}

\begin{figure}[H]
    \centering
    \includegraphics[width=0.8\linewidth]{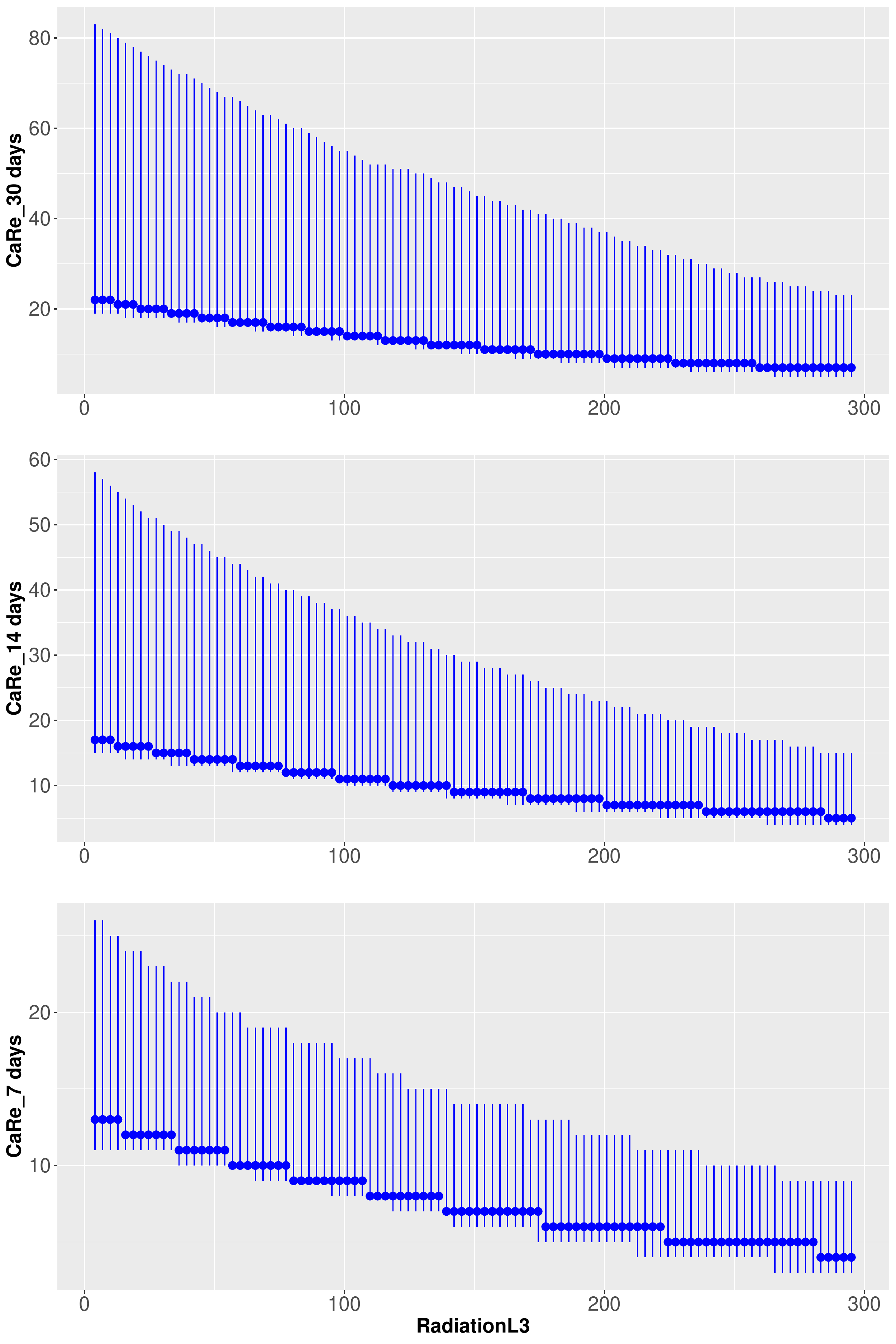}
    \caption{Positive cases: estimated CaRe for $h = 30, 14, 7$ days (panels from top to bottom) with respect to lagged radiation. The bars correspond to 95\% intervals. The other covariates are fixed to their mean values.}
    \label{fig:Qpositive_radiation}
\end{figure}

\begin{figure}[H]
    \centering
    \includegraphics[width=0.8\linewidth]{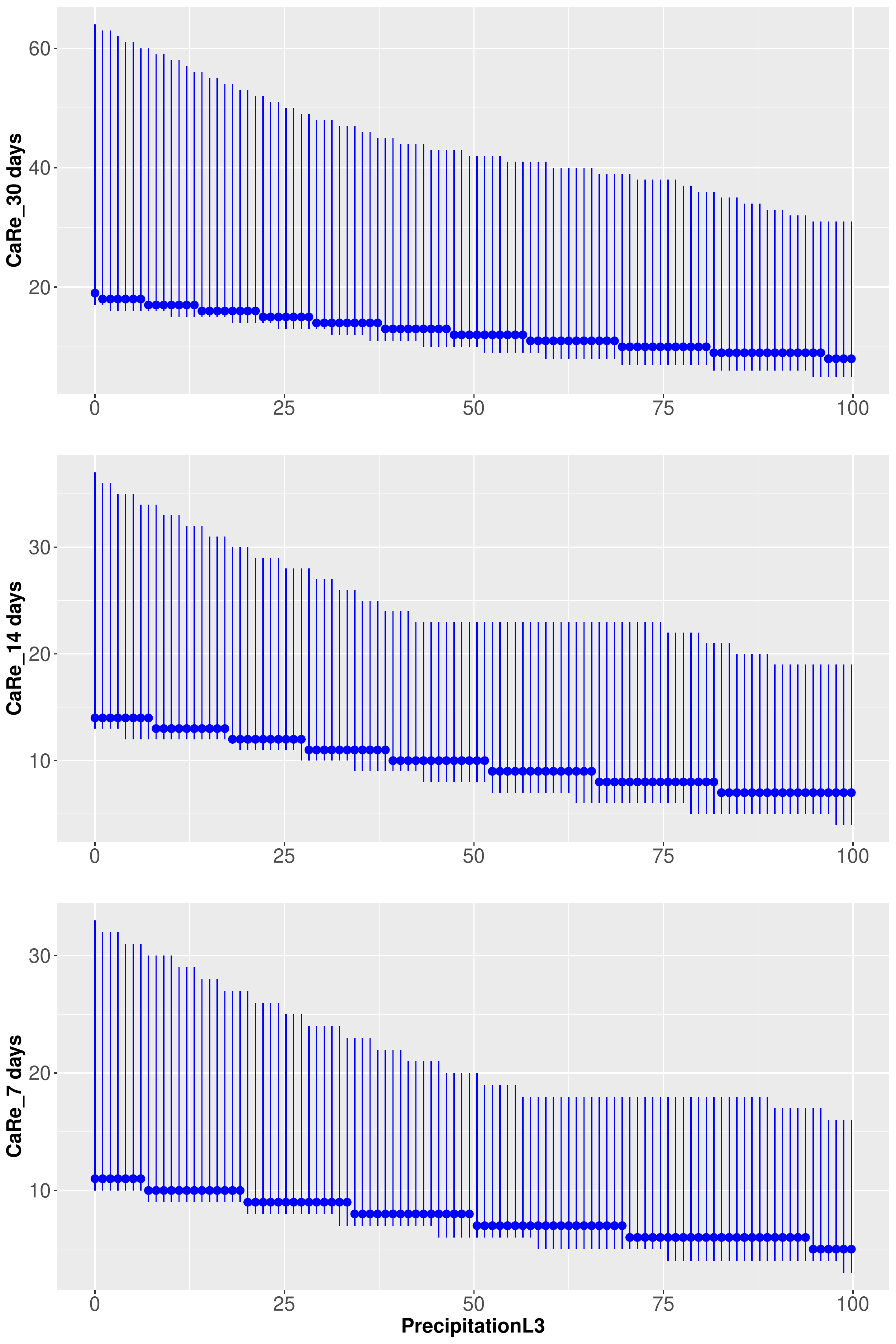}
    \caption{Positive cases: estimated CaRe for $h = 30, 14, 7$ days (panels from top to bottom) with respect to lagged precipitation. The bars correspond to 95\% intervals. The other covariates are fixed to their mean values.}
    \label{fig:Qpositive_precip}
\end{figure}

For the positive cases, the $p$\%-CaRe corresponds to the regime of congestion. Figures \ref{fig:Qpositive_mintemp}, \ref{fig:Qpositive_radiation} and \ref{fig:Qpositive_precip} show some $p$\%-CaRe estimated curves and respective intervals obtained via posterior simulation, based on different values of the meteorological factors. For instance, the bottom panel of Figure \ref{fig:Qpositive_mintemp} corresponds to the 0.86\%-CaRe, that is the number of positive cases that can be exceeded one time over a time horizon of 7 days. This point estimate decreases from 18, for a minimum temperature of -10 degC, to 11, for 0 degC. For a horizon of 7 days, the bottom left panel of Figure \ref{fig:Qpositive_radiation} shows that the number of positive cases that can be exceeded once a week decreases from 16, in the case of no sun, to 3, when radiation is at its highest value. This evidence should, however, be interpreted bearing in mind the widths of the intervals which are large.

Supplementary Figures \ref{fig:Qnegative_radiation} and \ref{fig:Qnegative_precip} show the CaRe estimated values with their simulated intervals as a function of meteorological predictors, for different time horizons for negative cases. These values convey information on the risk of inefficiency in the flu testing process, revealed by a very high number of negative cases. The risk of inefficiency considerably decreases both in sunny and raining periods. The top panel of Figure \ref{fig:Qnegative_precip} shows the estimated number of negative cases that can be exceeded once a month in terms of amount of precipitation. For a 30-day horizon, CaRe goes from 50, in the scenario of no rain, to 30 when it rains.

Large estimated $p$\%-CaRe values for the odds positives, as shown in Figure \ref{fig:Qodds_mintemp} with their simulated intervals, correspond to the epidemic regime. Note that the intervals are thinner as compared to those obtained for the discrete responses, but the values themselves are smaller (between 0 and 1). 

\subsection{Outliers detection}

\begin{figure}[H]
    \centering
    \includegraphics[width=0.9\linewidth]{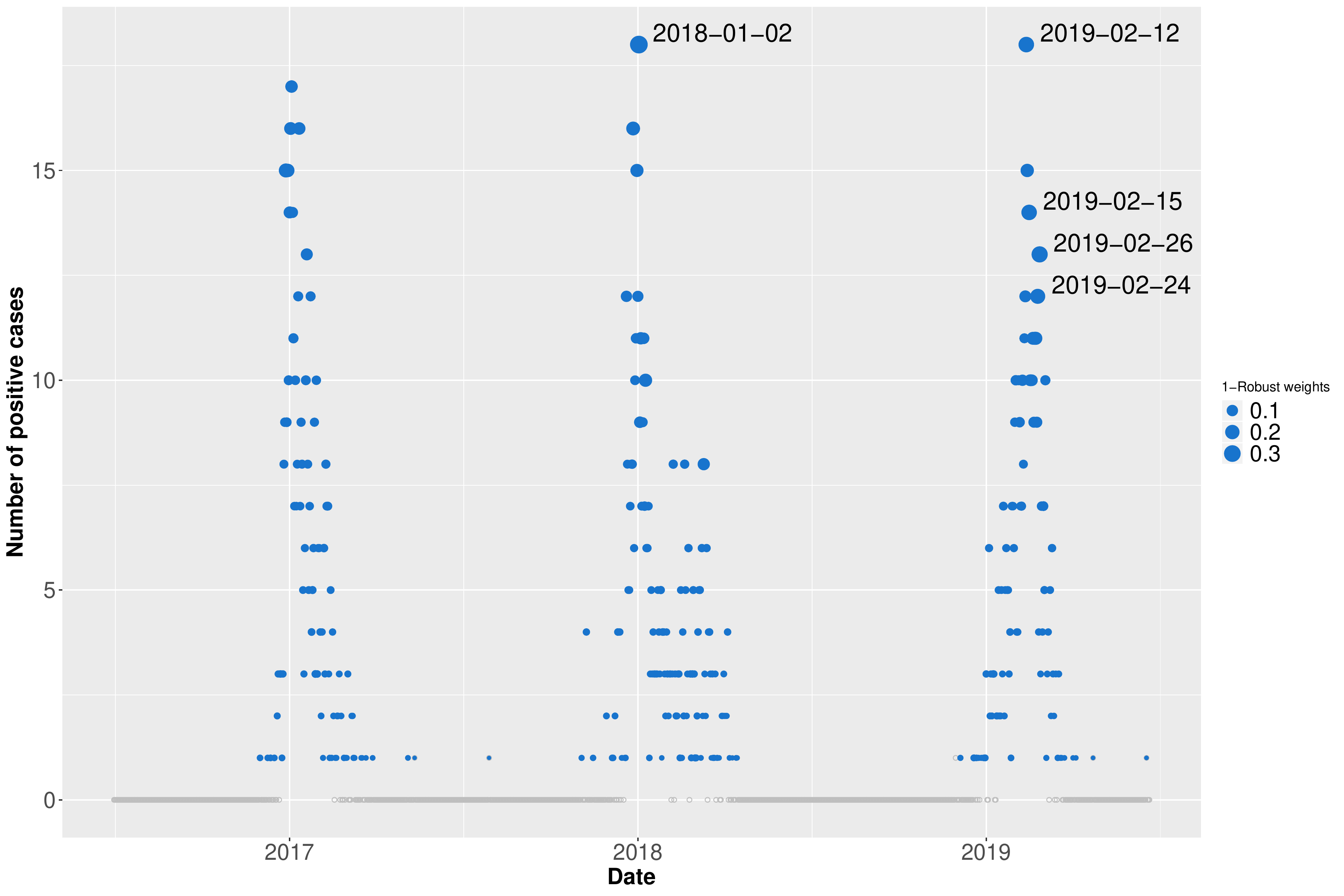}
    \caption{Robustness weights from the model fitted to positive cases.}
    \label{fig:outliers_pos}
\end{figure}

\begin{figure}[H]
    \centering
    \includegraphics[width=0.9\linewidth]{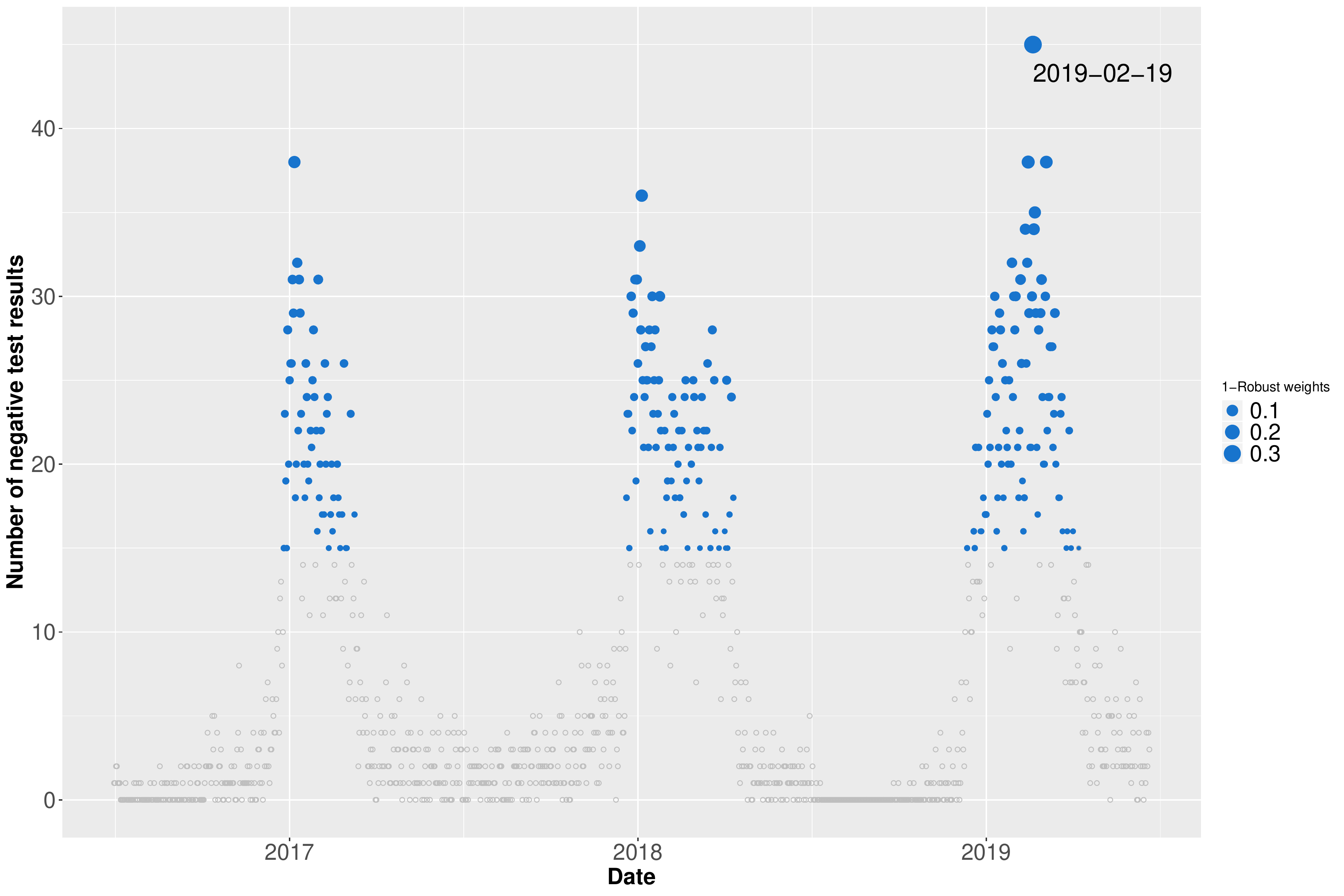}
     \caption{Robustness weights from the model fitted to negative cases.}
    \label{fig:outliers_neg}
\end{figure}

\begin{figure}[H]
    \centering
    \includegraphics[width=0.9\linewidth]{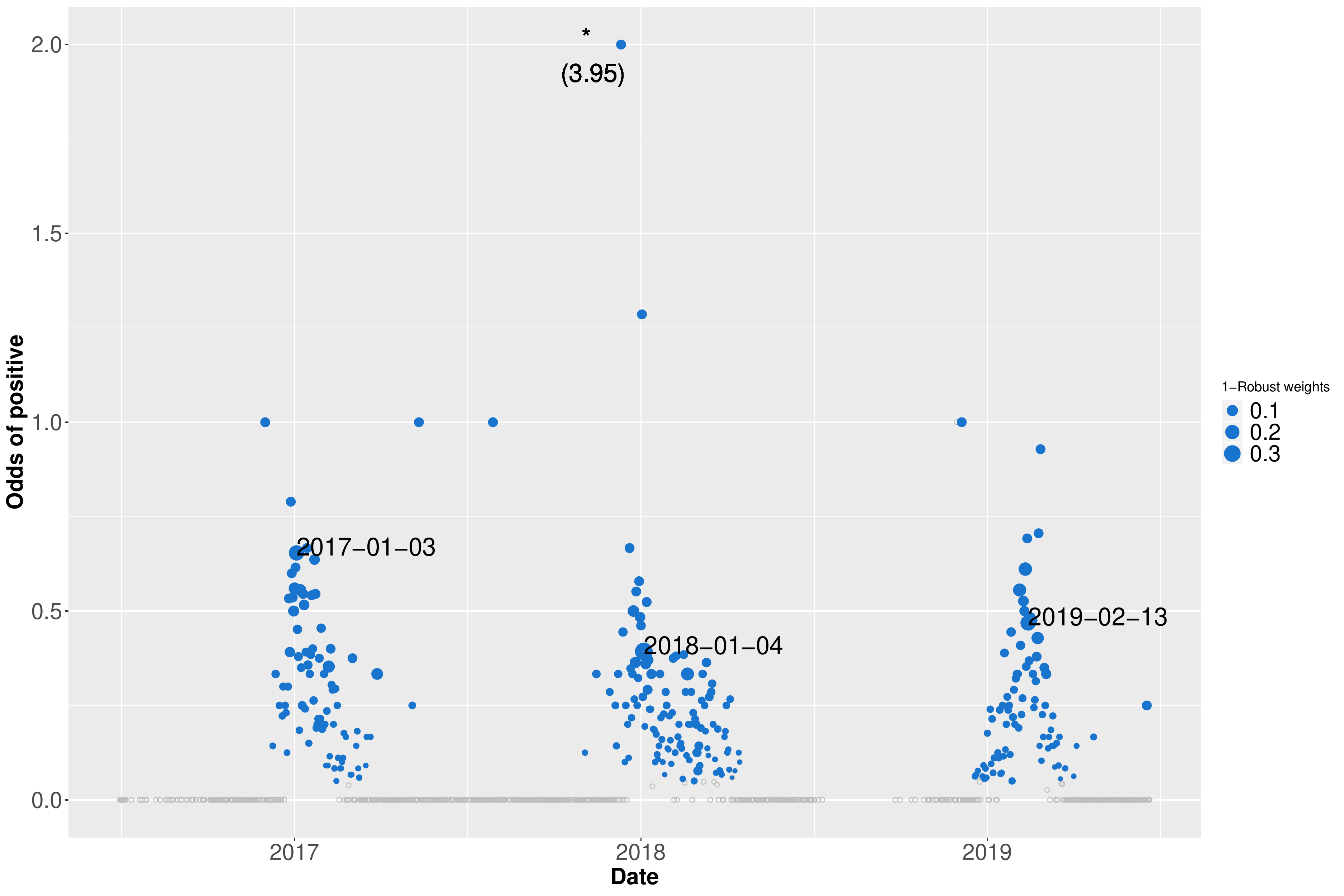}
     \caption{Robustness weights from the model fitted to odds of positive cases. The star (*) at the top of the graph indicates that an observation (odds = 3.95 on 08/11/2017 and robustness weight = 0.93) has not been represented.}
    \label{fig:outliers_odds}
\end{figure}

The hospital flu testing process depends on managerial and/or decision making instances and may lead to outlying records; our robust methodology is capable of detecting these abnormal values. Figures \ref{fig:outliers_pos}, \ref{fig:outliers_neg} and \ref{fig:outliers_odds} show the robustness weight $w = \rho_c^{\prime}$ (see Section~\ref{sct:gamlss}) for each observation. These are obtained as a by product of the parameter estimation process. The size of the circles is proportional to $(1-w)$. These weights can be used to identify outliers: the lower the weight, the more likely the observation is to be outlying. We identify the points with the smallest weights, accompanied by their date of occurrence, from the figures. The observations with the smallest weights are not systematically those with large observed values, and, conversely, the largest observations are not systematically considered outliers. This can be seen in particular for the odds of positive cases.

Dates such as 03/01/2017 and 02/01/2018 (which are after the Christmas break) are typically regarded as special days in the flu recording process. Days in February 2019 such as 24/02/2019 and 26/02/2019 in Figure \ref{fig:outliers_pos} relate to positive cases and correspond to school holidays. At school, children are super-spreaders. This is not the case during holidays when children are with their families. This and the fact that people are less tested during holidays contribute to a significant slow-down of the flu epidemic.

%% file: 05-simulation.ch
\label{sct:simulation}

To assess the empirical properties of our proposal in finite samples, we designed a simulation study inspired by our data analysis in Section~\ref{sct:models}. We will look at the quality of the estimated parameters and CaRe obtained using the proposed approach, both under the assumed (D-GPD or GPD) model and under contamination (i.e., in the presence of observations that deviate from the assumed model).

\subsection{D-GPD}
For D-GPD responses, we generated "clean" datasets from the model
\begin{equation}
    \label{simDGPDmodel}
\sqrt{\xi}= \alpha_{0} \ \ \ \mbox{and} \ \ \ 
\log(\sigma) = \beta_{0} + \beta_{1} \boldsymbol{x}_1 + \beta_{2} \boldsymbol{x}_2 + \beta_{3} \boldsymbol{x}_3,
\end{equation}
where the covariates distributions were chosen to mimic the behavior of \texttt{MintempL3}, \texttt{RadiationL3} and \texttt{PrecipitationL3} in the application. More precisely, we simulated the covariates as follows: $\boldsymbol{x}_1 \sim {\mathcal N}(2.3, 14)$,  $\boldsymbol{x}_2 \sim \Gamma(1.55, 0.02)$ and $\boldsymbol{x}_3 \sim \mbox{lognormal}(0.71, 3.12)$. These were kept fixed throughout the simulation replicates. The parameters, inspired by our case study, were set to $\alpha = 0.01$ (hence $\xi = 10^{-4}$), and $\boldsymbol{\beta} = (2,-0.05,-0.005,-0.01)^T$. Contaminated datasets were obtained by randomly setting $5\%$ of the response values to the maximum value observed in the sample. Again, this was inspired by the empirical application.

The sample size was set to $n=250$ (hence consistent with our application) and the number of replications to 500. We fitted models based on the robust and classical maximum likelihood estimators. For the robust approach, we chose $c=5.8$ to achieve a down-weighting of $0.95$.

\begin{figure} [H]
    \centering
    \includegraphics[width=0.8\linewidth]{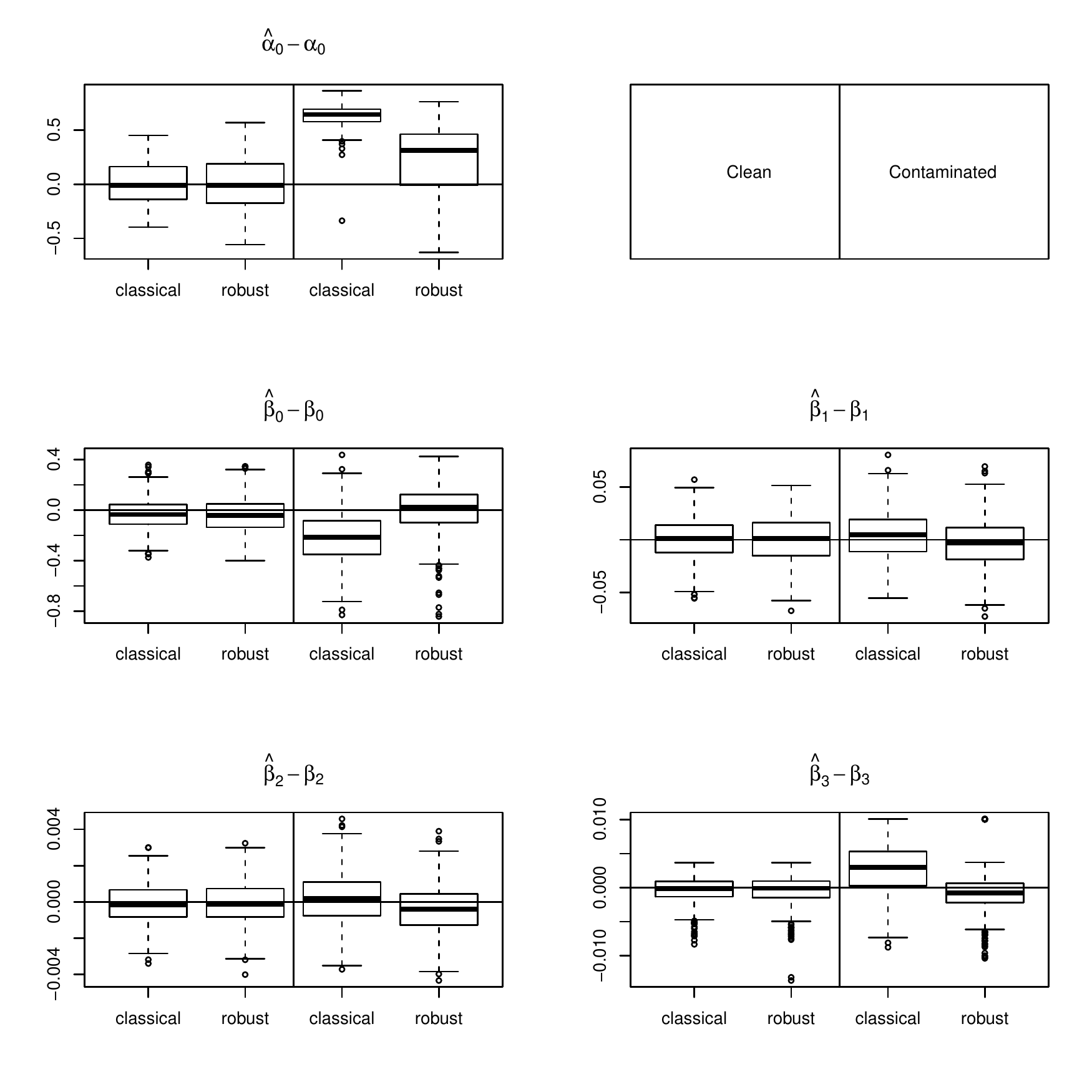}
        \caption{Boxplots of the centered parameter estimates for model~\eqref{simDGPDmodel}. In each panel, the left boxplots refer to the clean data setting, and the right ones to the contaminated data setting.}
    \label{fig:boxplotDGPD}
\end{figure}

Figure~\ref{fig:boxplotDGPD} shows the centered estimated parameter estimates for model~\eqref{simDGPDmodel}. By looking at the left boxplots of each panel, corresponding to the clean data setting, we see that both boxplots are centered around zero suggesting unbiased estimation of the model parameters. We also see that the variability of the robust estimator's estimates are slightly larger than the variability of those obtained using its classical counterpart; this is  expected because of the loss of efficiency of the robust estimator with respect to maximum likelihood. The story is different when looking at the right boxplots of each panel, corresponding to the contaminated data. All in all, the robust estimator performs much better than the classical one. For all the components of $\boldsymbol{\beta}$, except $\beta_2$, the robust estimator shows no bias. This is not the case for the classical estimator, which is influenced by the outliers, sometimes heavily, notably for $\beta_0$ and $\beta_3$. Surprisingly, the robust estimator shows some bias for $\beta_2$, but its magnitude is small. Also note that contamination affects the variability of the classical estimator, which is larger than that of the robust estimator for all the components of $\beta$.
The estimation of $\alpha_0$ under contamination seems more difficult. Both estimators show bias, even though that bias is much smaller for the robust estimator. It is worth recalling that a robust estimator guarantees that the bias under contamination does not explode, but it does not guarantee that it will vanish. Had we increased the amount of contamination or its strength, we would have expected the bias of the classical estimator to explode, but not that of the robust approach.

\begin{figure} [H]
    \centering
    \includegraphics[width=1.1\linewidth]{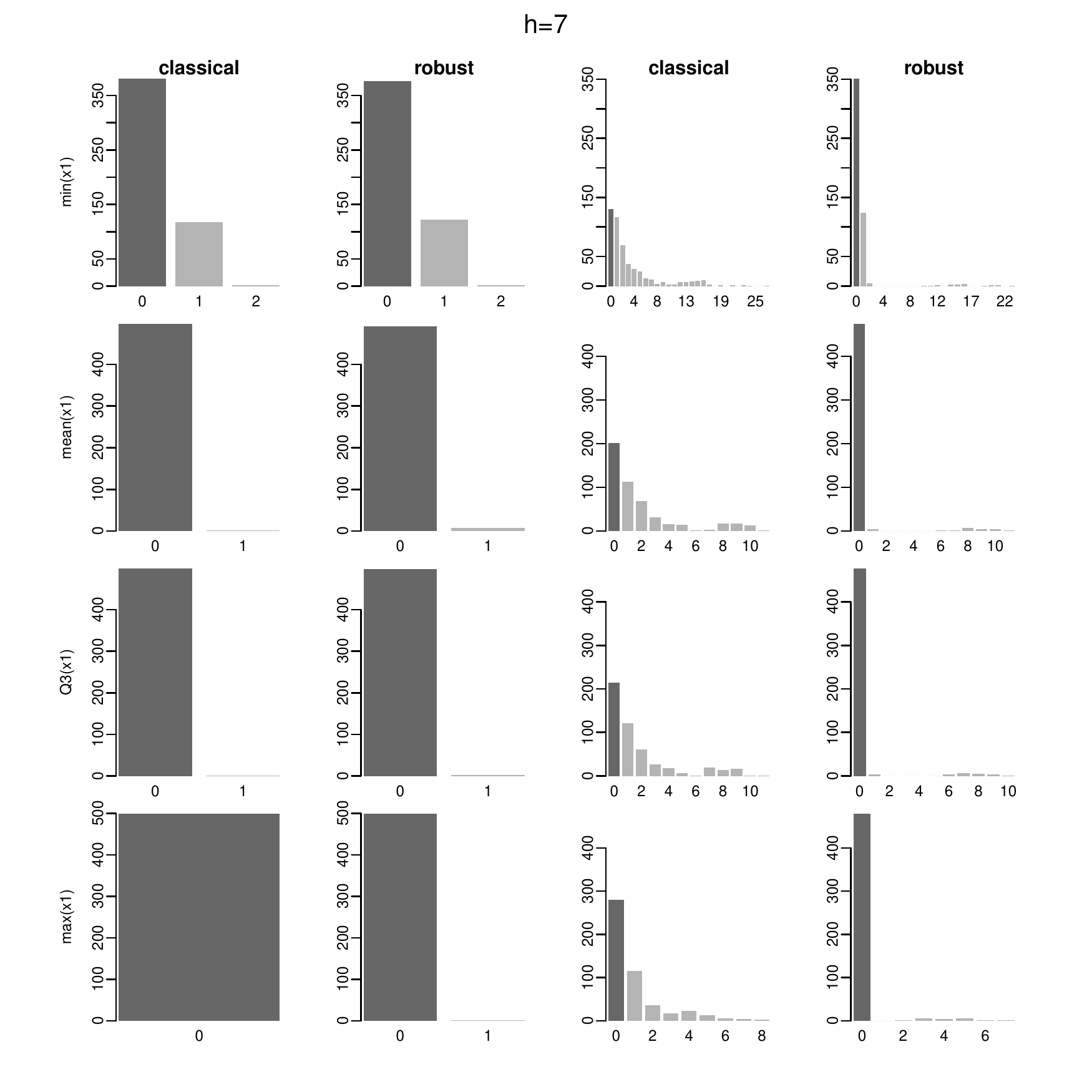}
        \caption{CaRe estimates for $h=7$ for four values of $x_{1}$: its minimum value (first row), its average (second row), its third quartile (third row) and its maximum (fourth row). The other covariates are fixed at their mean values. The true population value is identified by a darker bar. The first two columns correspond to the clean data setting, whereas the last two columns correspond to the contaminated data setting.}
    \label{fig:CaReDGPDh7x1}
\end{figure}

Figure~\ref{fig:CaReDGPDh7x1} reports the estimated CaRe for $h=7$ as a function of $x_{1}$, with $x_2$ and $x_3$ fixed at their mean values. CaRe is a discrete positive value (with relatively few different values in our simulation setting) and we depict it using barplots (histograms) for four representative values across the range of $x_1$: its minimum (first row), its average (second row), its third quartile (third row) and its maximum (fourth row). In each panel, the true population value is identified by the dark bar. The comparison of the first two columns confirms that the robust and classical estimators perform equally well and their estimates either match the population value or are just off by one unit. Looking at the results under contamination and comparing the last two columns of the figure, we observed that the robust estimator performs much better than the classical one in that it identifies the true population value more often (roughly $95\%$ of the times, except for $\min(x_1)$ where this is about $80\%$). In contrast, the classical CaRe estimator shows large variability and misses the population target quite often. The quality of the CaRe estimates is not uniform across the range of $x_{1}$, with lower values of the covariate being more challenging.

We provide in the supplementary material similar figures for all the other combinations of horizons $h$ and covariates; the conclusions do not change. Moreover, as expected, we observe that the performance of both estimators worsen as $h$ increases, and that the CaRe seems to be difficult to estimate for low values of $x_3$.

To sum up, our simulation study demonstrates that our distributional regression approach is effective in estimating models with D-GPD responses, and that the robust version of the estimator can successfully cope with contaminated data.

\subsection{GPD}
In this case, we generated data using the model
\begin{equation}
\label{simGPDmodel}
\log (\xi+ 0.5 )= \alpha_{0} \ \ \ \mbox{and} \ \ \ 
\log(\sigma) = \beta_{0} + \beta_{1} \boldsymbol{x}_1,
\end{equation}
where  $\boldsymbol{x}_1 \sim {\mathcal N}(2.3, 14)$, which reproduces what we see in our application for \texttt{MintempL3}. We consider $\alpha = -2 $ (hence $\xi = -0.36 )$ and $\boldsymbol{\beta} = (-1.3,-0.1)^T$. Contaminated datasets were obtained by randomly setting $5\%$ of the response values to the maximum value observed in the sample. The sample size was set to $n=250$ and the number of replications to 500. Here as well, we fitted models based on the robust and classical maximum likelihood estimators, and for the robust approach, we chose $c=2.3$ to achieve a downweighting of $0.95$.

\begin{figure} [H]
    \centering
    \includegraphics[width=0.8\linewidth]{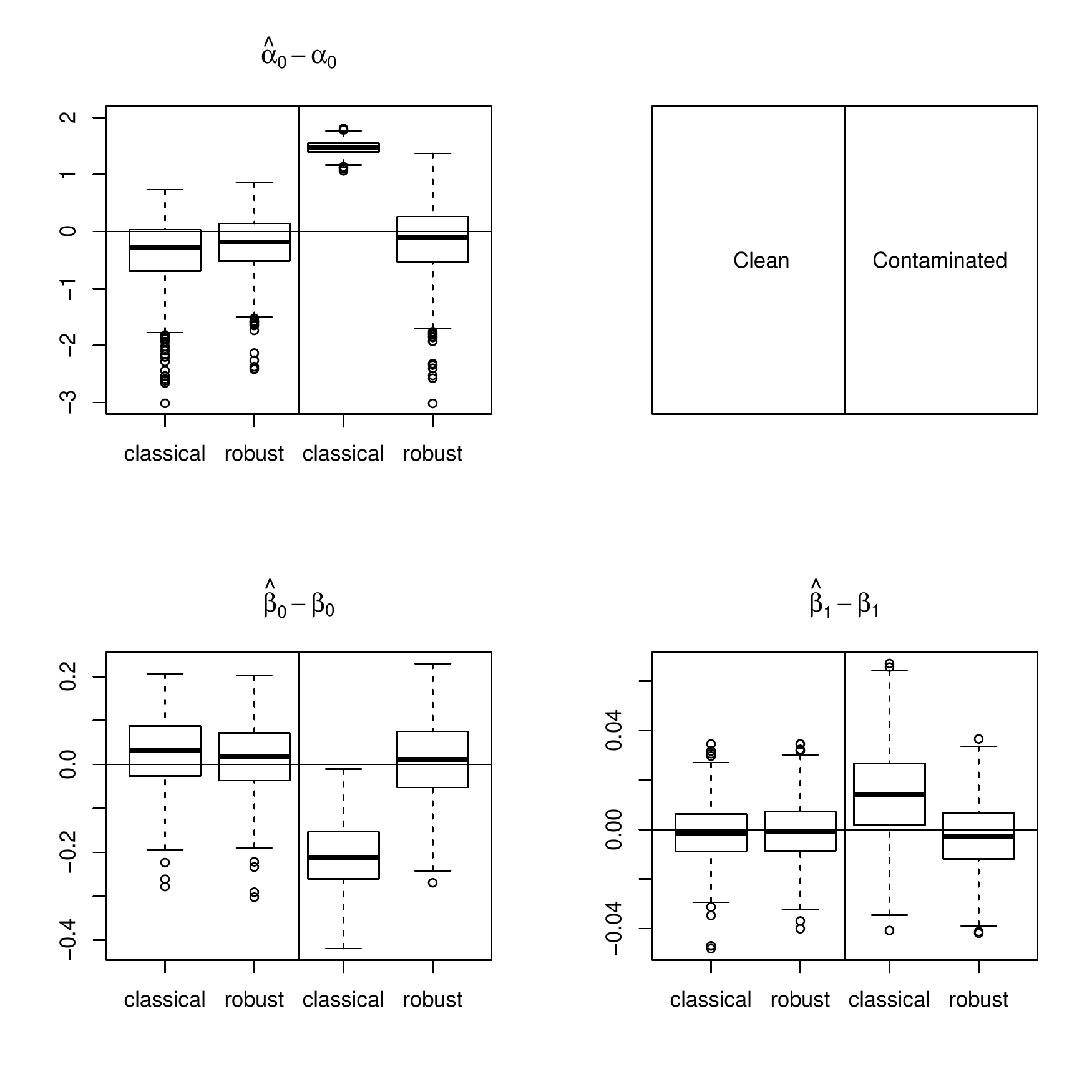}
        \caption{Boxplots of the centered parameter estimates of model~\eqref{simGPDmodel}. In each panel, the left boxplots are for the clean data setting, and the right boxplots for the contaminated data setting.}
    \label{fig:boxplotGPD}
\end{figure}

The analysis of the boxplots of the centered parameter estimates for model~\eqref{simGPDmodel} conveys the same conclusions as for the case of D-GPD: the robust estimator performs as well as its classical counterpart when using clean data, and outperforms the classical estimator under contamination. Furthermore, it is known that the GPD is difficult to fit and our findings suggest that our robust approach has a stabilizing property even for clean data where it corrects what seems to be a finite sample bias of the maximum likelihood estimator when estimating $\alpha_0$. It has been previously observed in the literature that robust estimators converge faster than their classical counterparts to their asymptotic distribution.

\begin{figure} [H]
    \centering
    \includegraphics[width=1.1\linewidth]{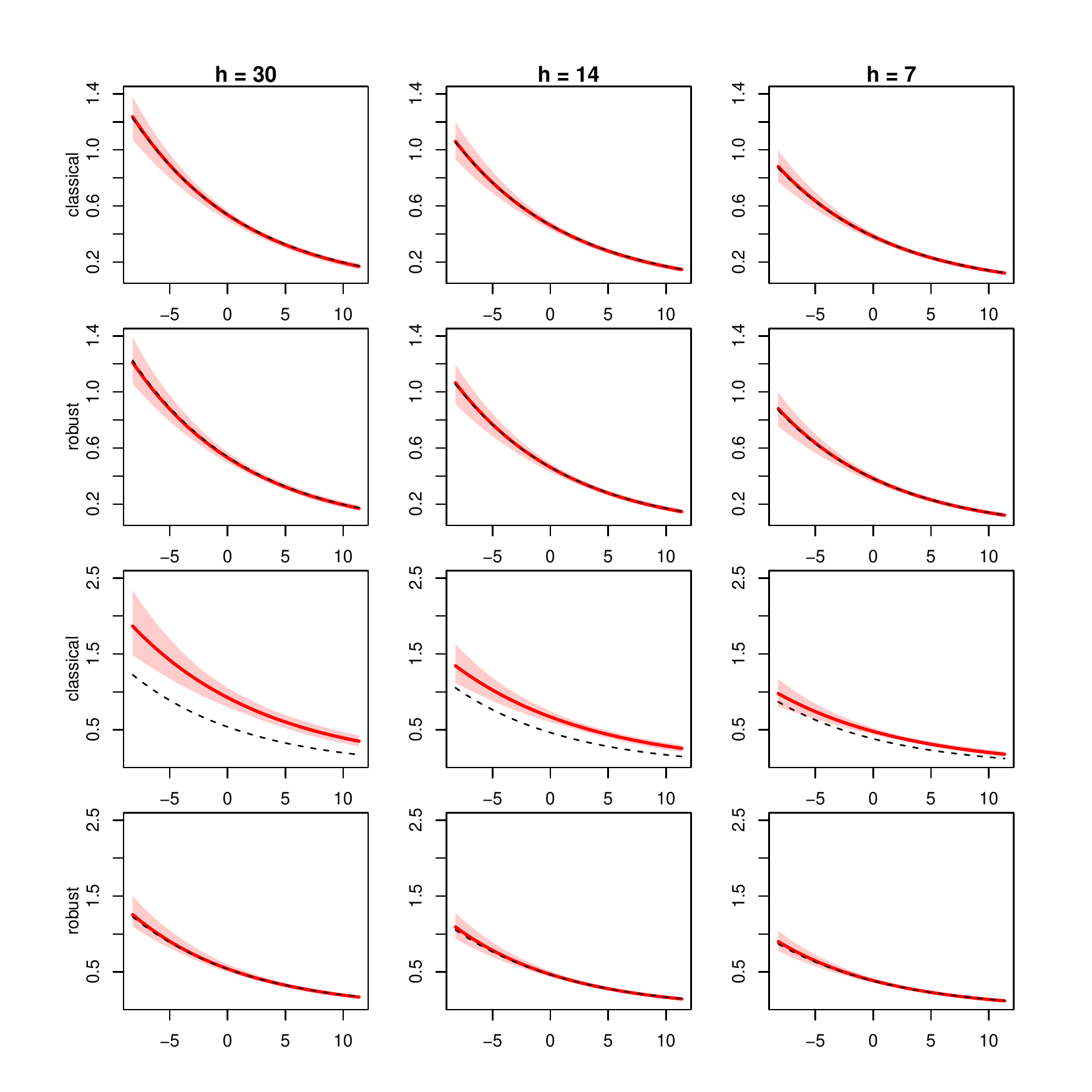}
        \caption{Functional boxplots of CaRe estimates as a function of $x_{1}$ for $h=30$ (first column), $h=14$ (second column) and $h=7$ (third column). The solid red line is the median curve and the envelope represents the $50\%$ deepest, or most central, observations. The black dashed line is the true population CaRe. The first two rows of panels are for the clean data setting, the last two rows are for the contaminated data setting. }
    \label{fig:CaReGPD}
\end{figure}

Figure~\ref{fig:CaReGPD} shows functional boxplots \citep{sun2011functional} for the CaRe estimates as a function of $x_{1}$. The solid red line is the median curve and the envelope represents the $50\%$ deepest, or most central, curves, much like the interquartile range in a regular boxplot. The black dashed line is the true population CaRe. The first two rows of the panels refer to the clean data setting, whereas the last two rows to the contaminated data setting.

From the two top rows, we observe that for clean data both the classical and robust estimates are aligned with the population values. When the data are contaminated (third and fourth rows of panels) the classical estimator completely misses the target by overestimating the CaRe, whereas the robust estimator continues to perform  well. The wider pink envelopes also suggest that the classical estimator has larger variability than the robust one under contamination. The unsatisfactory behavior of the classical estimator worsens with an increasing $h$, as one would expect.

In summary, once again, we observe that our distributional regression approach works well in estimating models with GPD responses, with the robust version being rather stable under contamination.

%% file: 05-conclusion.ch
\label{sct:conclusion}
Seasonal epidemics may lead to hospitals congestion. In this paper, we use extreme value theory to study the occurrence of large numbers of flu cases in a hospital. To this end, we developed and implemented in \texttt{GJRM} a robust regression-type methodology that allows for non-identically distributed discrete and continuous extremes, and that deals with outlying data. The response variables of interest (the positive and negative cases together with the odds of positives) are statistically explained by meteorological variables. Although the models selected for this case study are based on parametric covariate effects, our software implementation allows for very general non-parametric functional forms, which would most likely be required for larger datasets.  


Taking the point of view of the hospital, which needs to manage admission capacities, we introduced the notion of charge-at-risk whose estimation, based on meteorological factors, can serve as a quantitative tool to alert the hospital and allow time to prepare for a possible congestion. The introduced approach could be applied to several types of seasonal virus data such as those deriving from the new virus SARS-CoV-2. 




%% file: 06-appendix.ch
\section{Analytical Derivatives}\label{sct:appex}
This section provides the analytical score and Hessian components of the model's log-likelihood, for the GPD and D-GPD distributions. We also report the expectations and variances for the two distributions, and for the D-GPD we explain how variates are simulated.

\subsection{GPD}
	If $Y$ follows a GPD distribution with shape parameter $\xi$ and scale parameter $\sigma$, then its cumulative distribution function is $F_{(\sigma,\xi)}(y) = 1 - G_{(\sigma,\xi)}(y)$ (with $G_{(\sigma,\xi)}(y)$ given in \eqref{gpd_asymptotic}) which is defined for $\sigma>0$ and for $\{y: y>0 \quad \text{and} \quad  1+\frac{\xi y}{\sigma}>0  \}$. 
	
	Then the probability density function is
	$$ 	g_{(\sigma,\xi)}(y)= \frac{1}{\sigma}\left(1+\frac{\xi y}{\sigma}\right)^{(-1/\xi-1)}.$$	
	
	The expectation of $Y$ is $E(Y)= \frac{\sigma}{1-\xi}$ and the variance is $Var(Y)=\frac{\sigma^2}{(1+\xi)^2(1-2\xi)}$.
	
	Having observed $\mathbf{y}=\{y_{1},y_{2}, \cdots, y_{n}\}$ as independent realizations of $Y$ the log-likelihood is

	$$l(\mathbf{y};\xi,\sigma)= \log \left( \prod_{i=1}^{n} g_{(\sigma,\xi)}(y_i) \right)=  -n\log(\sigma)-\left(1+1/\xi\right)\sum_{i=1}^{n} \log\left(1+\frac{\xi y_{i}}{\sigma}\right).$$
	
	The first and second derivatives with respect to $\xi$ and $\sigma$ are
	$$l_{\xi}= \frac{1}{\xi^2}\sum_{i=1}^{n}\log \left(1+\frac{\xi y_{i}}{\sigma}\right)-(1+1/\xi)\sum_{i=1}^{n}\frac{y_{i}/\sigma}{\left(1+\frac{\xi y_{i}}{\sigma}\right)},$$
	
	$$l_{\sigma}= - \frac{n}{\sigma}+(1+1/\xi)\sum_{i=1}^{n} \frac{\frac{\xi y_{i}}{\sigma^2}}{\left(1+\frac{\xi y_{i}}{\sigma}\right)}= \sum_{i=1}^{n} \frac{y_{i}-\sigma}{\sigma(\sigma+\xi y_{i})},$$
	
	$$ l_{\xi \xi}= \frac{-2}{\xi^3}\sum_{i=1}^{n}\log\left(1+\frac{\xi y_{i}}{\sigma}\right)+\frac{2}{\xi^2}\sum_{i=1}^{n}\frac{y_{i}}{\sigma}\frac{1}{\left(1+\frac{\xi y_{i}}{\sigma}\right)}+(1+1/\xi) \sum_{i=1}^{n} \left(\frac{y_{i}}{\sigma}\right)^2\frac{1}{\left(1+\frac{\xi y_{i}}{\sigma}\right)^2}, $$
	
	$$l_{\sigma \sigma}= \frac{n}{\sigma^2}+ \left(1+1/\xi\right)\sum_{i=1}^{n}\frac{\frac{-2\xi y_{i}}{\sigma^3}-\frac{\xi y_{i}^2}{\sigma^4}}{\left(1+\frac{\xi y_{i}}{\sigma}\right)^2}= \sum_{i=1}^{n}\frac{\sigma^2 - \xi y_{i}^2-2\sigma y_{i}}{\sigma^2(\sigma+\xi y_{i})^2},$$
	
	$$l_{\sigma \xi} = - \frac{1}{\xi}\sum_{i=1}^{n}\frac{y_{i}/\sigma^2}{\left(1+\frac{\xi y_{i}}{\sigma}\right)}+(1+1/\xi)\sum_{i=1}^{n} \left(1+\frac{\xi y_{i}}{\sigma}\right)^{-2}= \sum_{i=1}^{n} \frac{y_{i}(y_{i}-\sigma)}{\sigma (\sigma+\xi y_{i})^2}.$$

	\subsection{D-GDP}
	
	If $Y$ follows a D-GPD distribution with shape parameter $\xi$ and scale parameter $\sigma$ then the probability mass function of $Y=r$, where $r \in N_{0}=\{0,1,2,...\}$ is, using Equation~\eqref{dgpd}, 
	$$ 
	DG_{(\sigma,\xi)}(r)= \left(1+\frac{\xi r}{\sigma}\right)^{(-1/\xi)}-\left(1+\frac{\xi(1+r)}{\sigma}\right)^{(-1/\xi)},
	$$
	which is defined for $\sigma >0$ and for $\{r: 1+\frac{\xi r}{\sigma}>0 \qquad \text{and} \qquad 1+\frac{\xi (r+1)}{\sigma}>0 \}$. 
	
	The mean and variance have to be calculated numerically. Specifically,
	
	$$E(Y)=\sum_{r=1}^{\infty}\left(1+\frac{\xi r}{\sigma}\right)_{+}^{-1/\xi}, \xi>0$$
	 and 
	 $$Var(Y)=E(Y^{2})-[E(Y)]^2  = \sum_{r=1}^{\infty}(2r-1)\left(1+\frac{\xi r}{\sigma}\right)_{+}^{-1/\xi} - \left[ \sum_{r=1}^{\infty}\left(1+\frac{\xi r}{\sigma}\right)_{+}^{-1/\xi}  \right]^2.$$ 
	 

	Having observed $\mathbf{y}=\{y_{1},y_{2}, \ldots, y_{n}\}$ as independent realizations of $Y$, the log-likelihood is
	$$l(\mathbf{y};\xi,\sigma)=\sum_{i=1}^{n} \log(p_{i}),$$
	where
	$$p_{i}=\left(1+\frac{\xi y_{i}}{\sigma}\right)^{(-1/\xi)}-\left(1+\frac{\xi(1+y_{i})}{\sigma}\right)^{(-1/\xi)}.$$

	 The first and second derivatives with respect to $\xi$ and $\sigma$ are
	 $$ l_{\xi}=\frac{\partial l}{\partial \xi}= \sum_{i=1}^{n}\frac{\frac{\partial p_{i}}{\partial \xi}}{p_i},$$
	 $$l_{\sigma}= \frac{\partial l}{\partial \sigma}= \sum_{i=1}^{n}\frac{\frac{\partial p_{i}}{\partial \sigma}}{p_i},$$
	 $$ l_{\xi \xi}=\frac{\partial^2 l}{\partial \xi^2}=\sum_{i=1}^{n} \frac{\frac{\partial^2 p_{i}}{\partial\xi^2} p_{i}-\left(\frac{\partial p_{i}}{\partial \xi}\right)^2}{p_{i}^2},$$
	 
	 $$ l_{\sigma \sigma}=\frac{\partial^2 l}{\partial \sigma^2}=\sum_{i=1}^{n} \frac{\frac{\partial^2 p_{i}}{\partial\sigma^2}.p_{i}-\left(\frac{\partial p_{i}}{\partial \sigma}\right)^2}{p_{i}^2},$$
	 
	 $$ l_{\sigma \xi}=\frac{\partial^2 l}{\partial \sigma \partial \xi}= \sum_{i=1}^{n}\frac{\frac{\partial^2 p_{i}}{\partial  \sigma \partial \xi} p_{i}- \left(\frac{\partial p_{i}}{\partial \sigma}\right)\left(\frac{\partial p_{i}}{\partial \xi}\right)}{p_{i}^2}, $$
	 
	 the elements of which are 
	 \begin{align*}
	 	\frac{\partial p_{i}}{\partial \xi} &= \left(1+\frac{\xi y_{i}}{\sigma}\right)^{-1/\xi}\left[-\frac{y_{i}}{\sigma\xi (1+\frac{\xi y_{i}}{\sigma})}+\frac{\log (1+\frac{\xi y_{i}}{\sigma})}{\xi^2}\right]\\ 
	 	& -\left(1+\frac{\xi (y_{i}+1)}{\sigma}\right)^{-1/\xi}\left[-\frac{(y_{i}+1)}{\sigma\xi (1+\frac{\xi (y_{i}+1)}{\sigma})}+\frac{\log (1+\frac{\xi (y_{i}+1)}{\sigma})}{\xi^2}\right],
	\end{align*}
	
	\begin{align*}
\frac{\partial p_{i}}{\partial \sigma} &= \frac{y_{i}}{\sigma^2}\left(1+\frac{\xi y_{i}}{\sigma}\right)^{-1/\xi -1}
 - \frac{(y_{i}+1)}{\sigma^2}\left(1+\frac{\xi (y_{i}+1)}{\sigma}\right)^{-1/\xi -1},
	\end{align*}
	
	\begin{align*}
	\frac{\partial^2 p_{i}}{\partial \xi^2} &=
	\left(\frac{\log \left(\frac{\left(y_{i}+1\right)\xi}{\sigma}+1\right)}{\xi^2}-\frac{y_{i}+1}{\sigma \xi \left(\frac{\left(y_{i}+1\right)\xi}{\sigma}+1\right)}\right)^2\left(\frac{\left(y_{i}+1\right)\xi}{\sigma}+1\right)^\frac{-1}{\xi}\\
	&+\left(-\frac{2\log \left(\frac{\left(y_{i}+1\right)\xi}{\sigma}+1\right)}{\xi^3}+\frac{2\left(y_{i}+1\right)}{\sigma\xi^2\left(\frac{\left(y_{i}+1\right)\xi}{\sigma}+1\right)}+\frac{\left(y_{i}+1\right)^2}{\sigma^2\xi\left(\frac{\left(y_{i}+1\right)\xi}{\sigma}+1\right)^2}\right)\left(\frac{\left(y_{i}+1\right)\xi}{\sigma}+1\right)^\frac{-1}{\xi}\\ & +\left(\frac{\log \left(\frac{y_{i}\xi}{\sigma}+1\right)}{\xi^2}-\frac{y_{i}}{\sigma\xi\left(\frac{y_{i}\xi}{\sigma}+1\right)}\right)^2\left(\frac{y_{i}\xi}{\sigma}+1\right)^\frac{-1}{\xi}\\
	& +\left(-\frac{2\log \left(\frac{y_{i}\xi}{\sigma}+1\right)}{\xi^3}+\frac{2y_{i}}{\sigma\xi^2\left(\frac{y_{i}\xi}{\sigma}+1\right)}+\frac{y_{i}^2}{\sigma^2\xi\left(\frac{y_{i}\xi}{\sigma}+1\right)^2}\right)\left(\frac{y_{i}\xi}{\sigma}+1\right)^\frac{-1}{\xi},
	\end{align*}
	
	\begin{align*}
	\frac{\partial^2 p_{i}}{\partial \sigma^2} &= \frac{y_{i}^2}{\sigma^4}(1+\xi)\left(1+\frac{\xi y_{i}}{\sigma}\right)^{(-1/\xi-2)} - \frac{2 y_{i}}{\sigma^3}\left(1+\frac{\xi y_{i}}{\sigma}\right)^{-1/\xi -1}\\
	&- \frac{(y_{i}+1)^2}{\sigma^4}(1+\xi)\left(1+\frac{\xi (y_{i}+1)}{\sigma}\right)^{(-1/\xi-2)} + \frac{2 (y_{i}+1)}{\sigma^3}\left(1+\frac{\xi (y_{i}+1)}{\sigma}\right)^{-1/\xi -1},
	\end{align*}
	
	\begin{align*}
	&	\frac{\partial^2 p_{i}}{\partial \sigma \partial \xi} = \frac{y_{i}}{\sigma^2}\left(1+\frac{\xi y_{i}}{\sigma}\right)^{(-1/\xi-1)}\left[ (-1/\xi -1) (\frac{y_{i}}{\sigma})\left(1+\frac{\xi y_{i}}{\sigma}\right)^{-1}+ \log \left(1+\frac{\xi y_{i}}{\sigma}\right)(\frac{1}{\xi^2})\right]\\
	&+	\frac{-(y_{i}-1)}{\sigma^2}\left(1+\frac{\xi (y_{i}-1)}{\sigma}\right)^{(-1/\xi-1)}
	\\
	&
	\left[ (-1/\xi -1) (\frac{(y_{i}-1)}{\sigma})\left(1+\frac{\xi (y_{i}-1)}{\sigma}\right)^{-1}+ \log \left(1+\frac{\xi (y_{i}-1)}{\sigma}\right)(\frac{1}{\xi^2})\right].
	\end{align*}

%% file: Supp.ch
\section{Models graphical results}

\begin{figure}[H]
    \centering
    \includegraphics[width=0.8\linewidth]{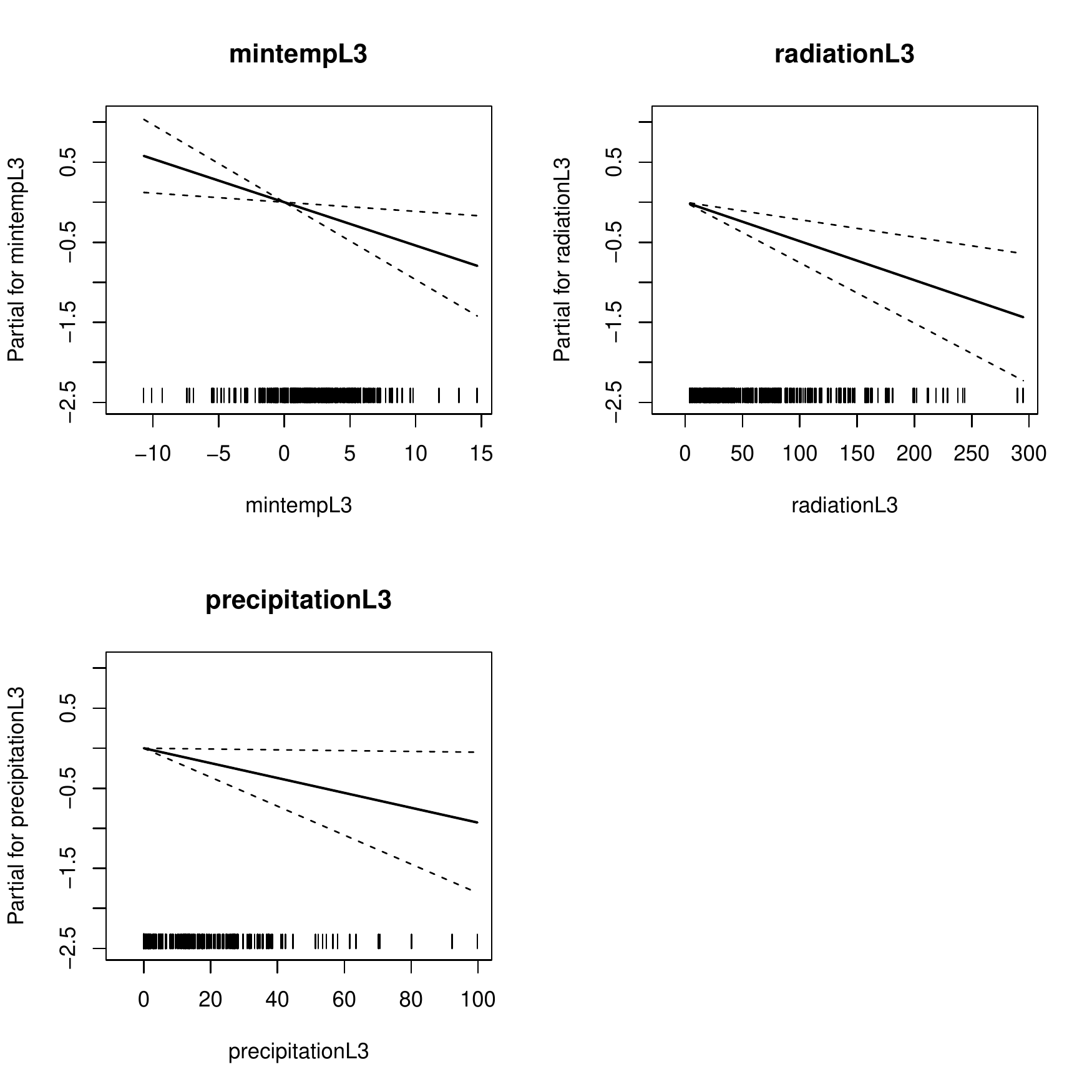}
    \caption{Covariate effects on the scale parameter $\sigma$ of the D-GPD for positive cases, and 95\% intervals.}
    \label{fig:positive-eq2}
\end{figure}

\begin{figure}[H]
    \centering
    \includegraphics[width=0.8\linewidth]{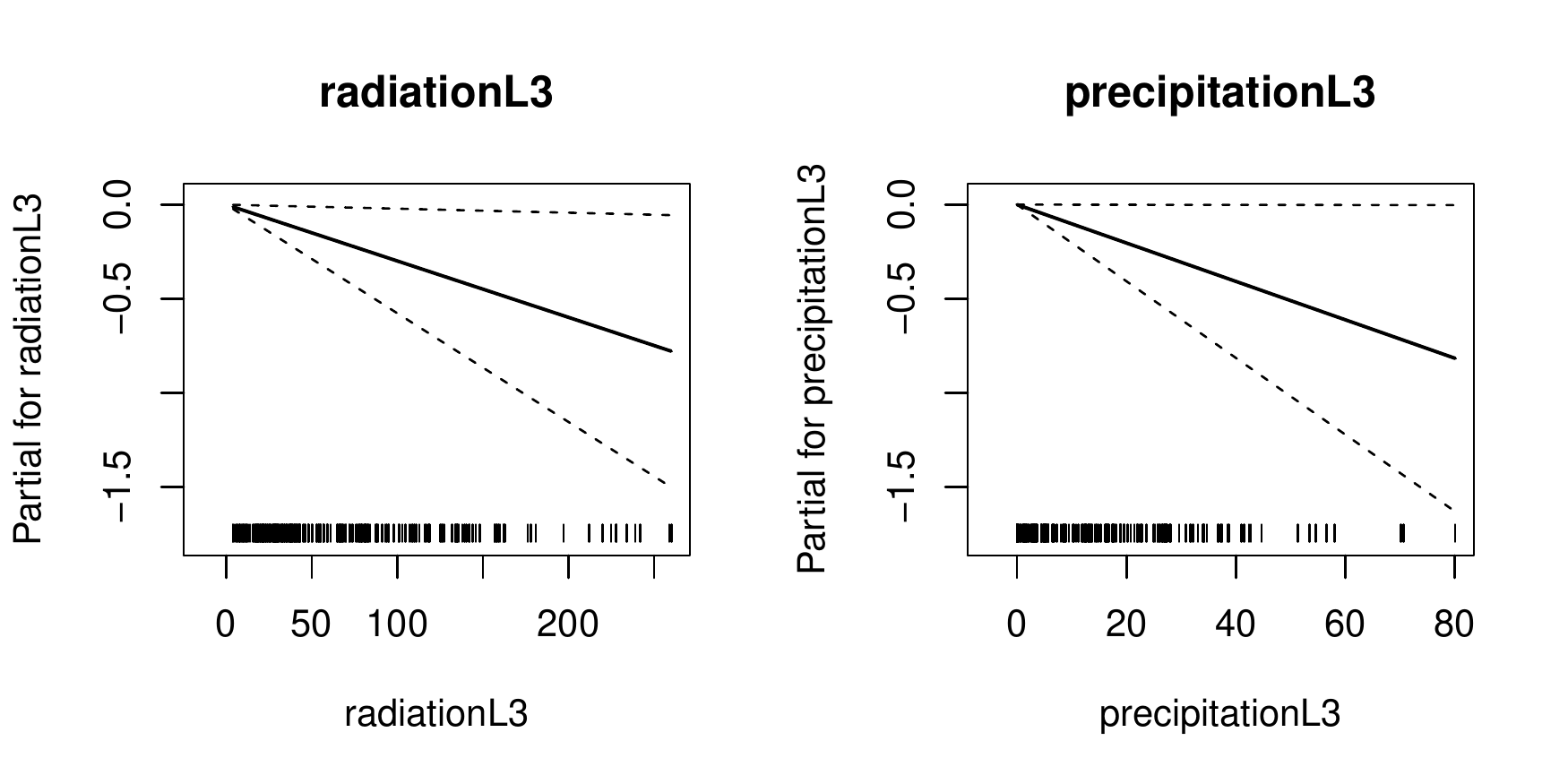}
        \caption{Covariate effects on the scale parameter $\sigma$ of the D-GPD for negative cases, and 95\% intervals.}

    \label{fig:negative-eq2}
\end{figure}

\begin{figure}[H]
    \centering
    \includegraphics[width=0.8\linewidth]{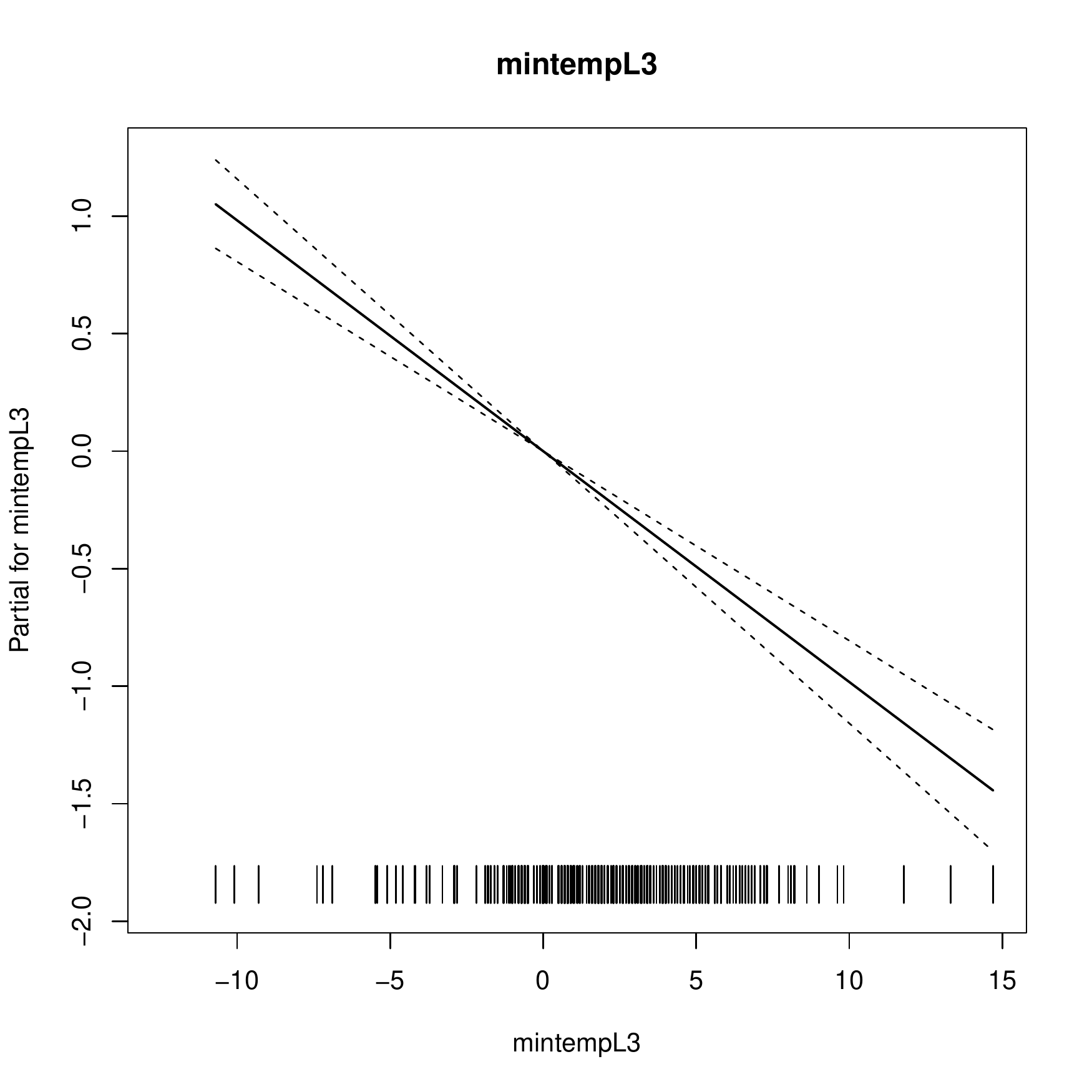}
    \caption{Covariate effect on the scale parameter $\sigma$ of the GPD for odds of positive cases, and 95\% intervals.}
    \label{fig:odds-eq2}
\end{figure}

 \begin{figure}[H]
    \centering
    \includegraphics[width=0.8\linewidth]{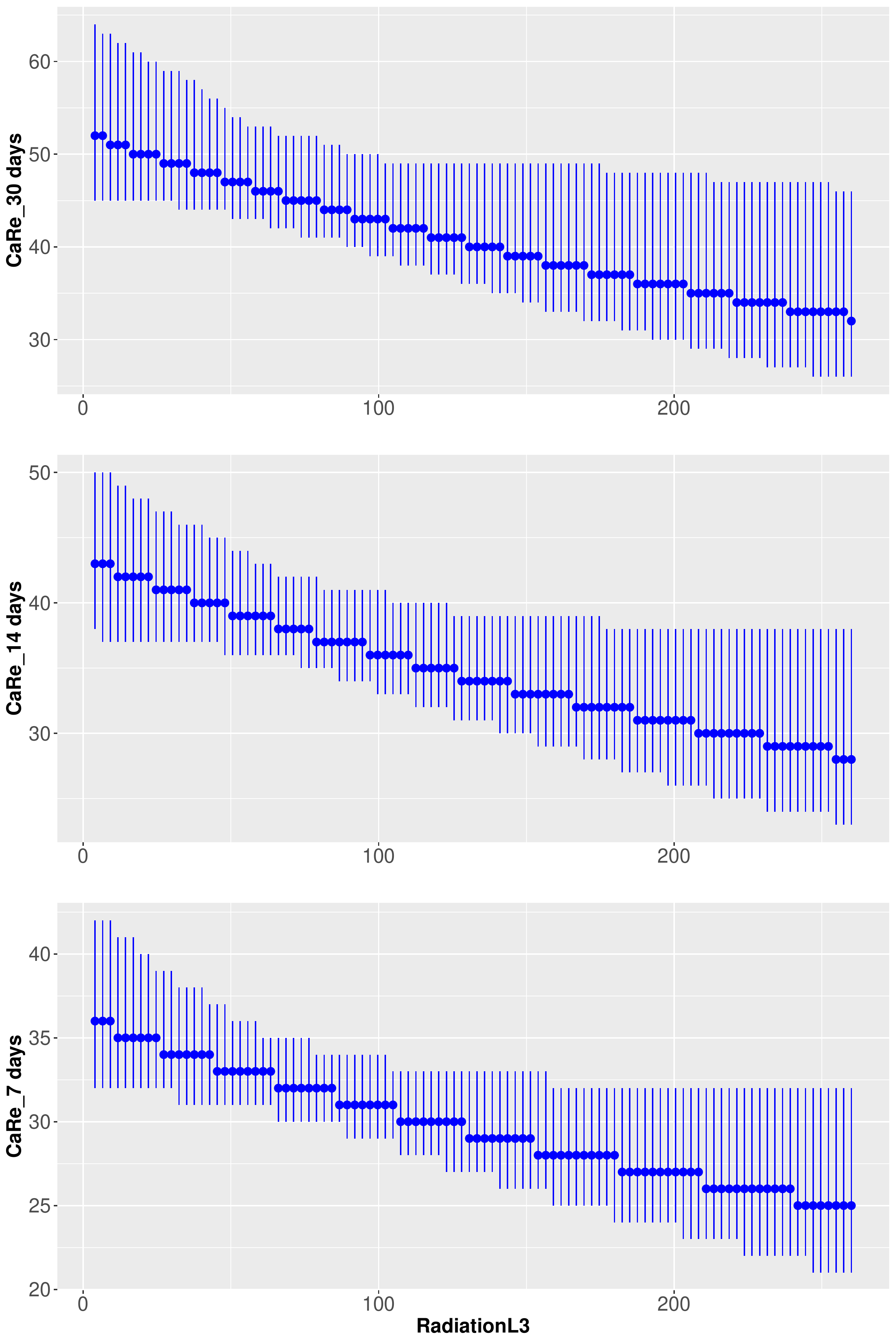}
    \caption{Negative cases: estimated CaRe for $h = 30, 14, 7$ days (panels from top to bottom) with respect to lagged radiation. The bars correspond to 95\% intervals. The other covariates are fixed to their mean values.}
    \label{fig:Qnegative_radiation}
\end{figure}

\begin{figure}[H]
    \centering
    \includegraphics[width=0.8\linewidth]{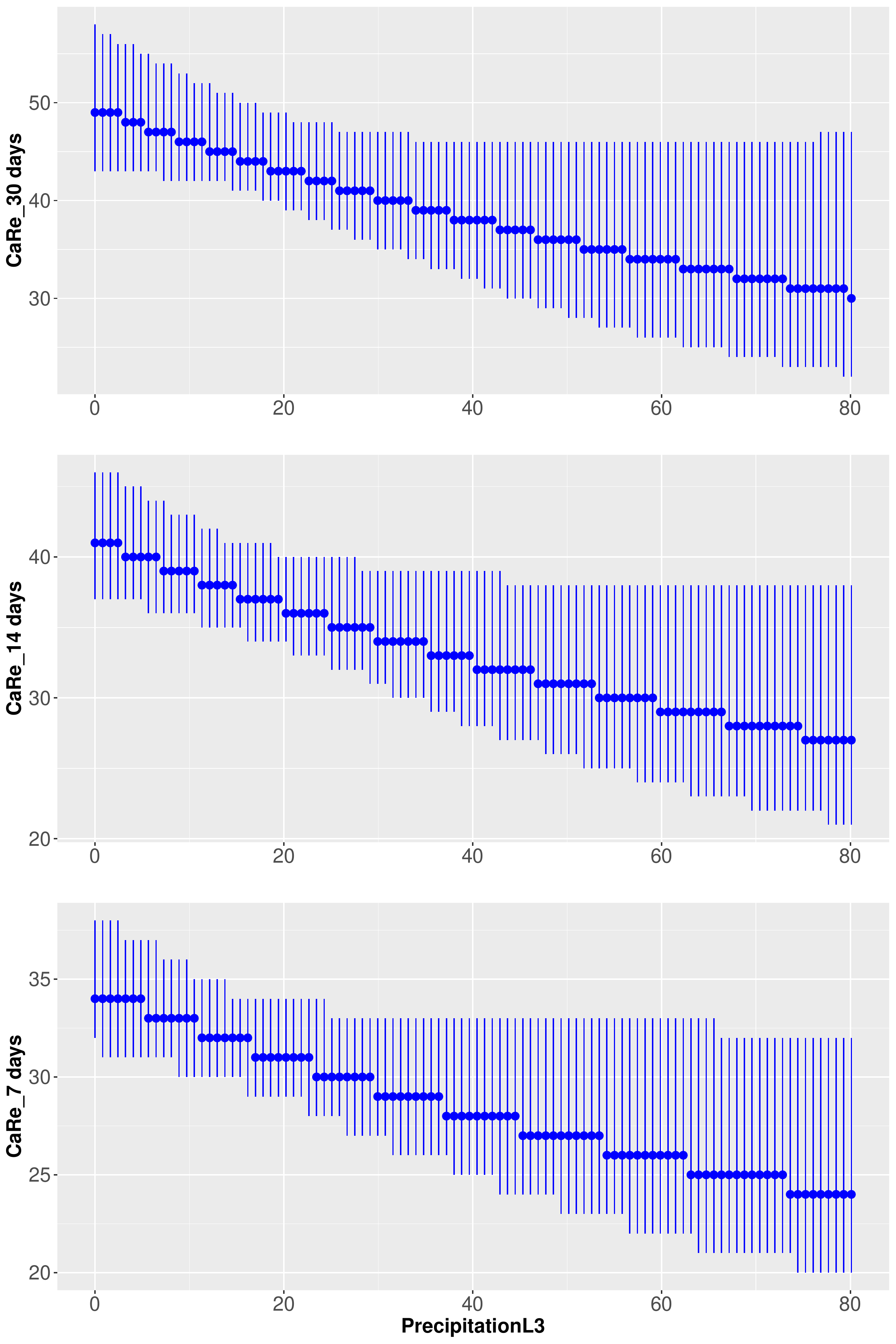}
    \caption{Negative cases: estimated CaRe for $h = 30, 14, 7$ days (panels from top to bottom) with respect to lagged precipitation. The bars correspond to 95\% intervals. The other covariates are fixed to their mean values.}
    \label{fig:Qnegative_precip}
\end{figure}

\begin{figure}[H]
    \centering
    \includegraphics[width=0.8\linewidth]{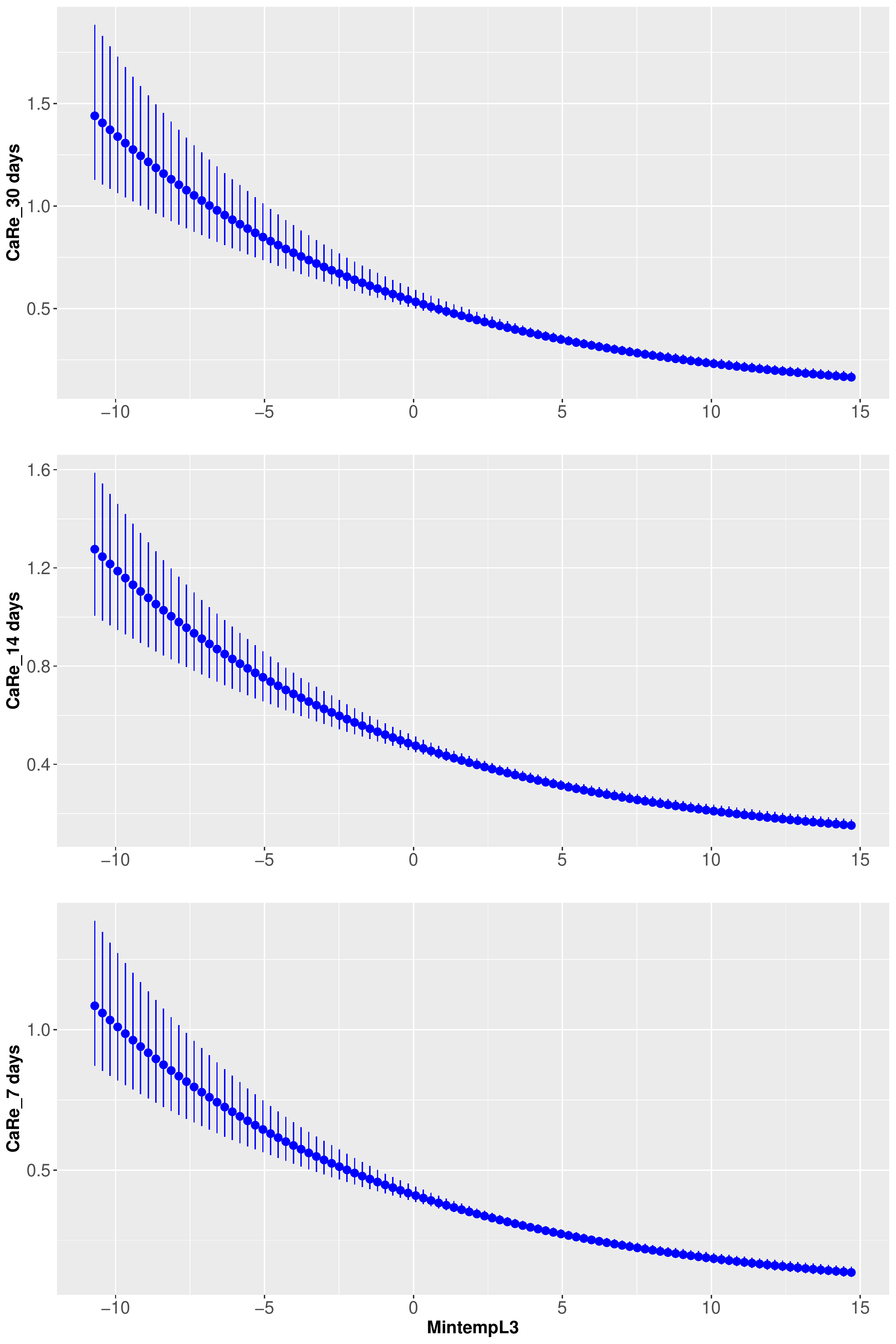}
    \caption{Odds of positive cases: estimated CaRe for $h = 30, 14, 7$ days (panels from top to bottom) with respect to lagged minimum temperature. The bars correspond to 95\% intervals.}
    \label{fig:Qodds_mintemp}
\end{figure}

%% file: 08-simulation-figures.ch
\section{Additional simulations results for D-GPD}

\begin{figure} [H]
    \centering
    \includegraphics[width=1.1\linewidth]{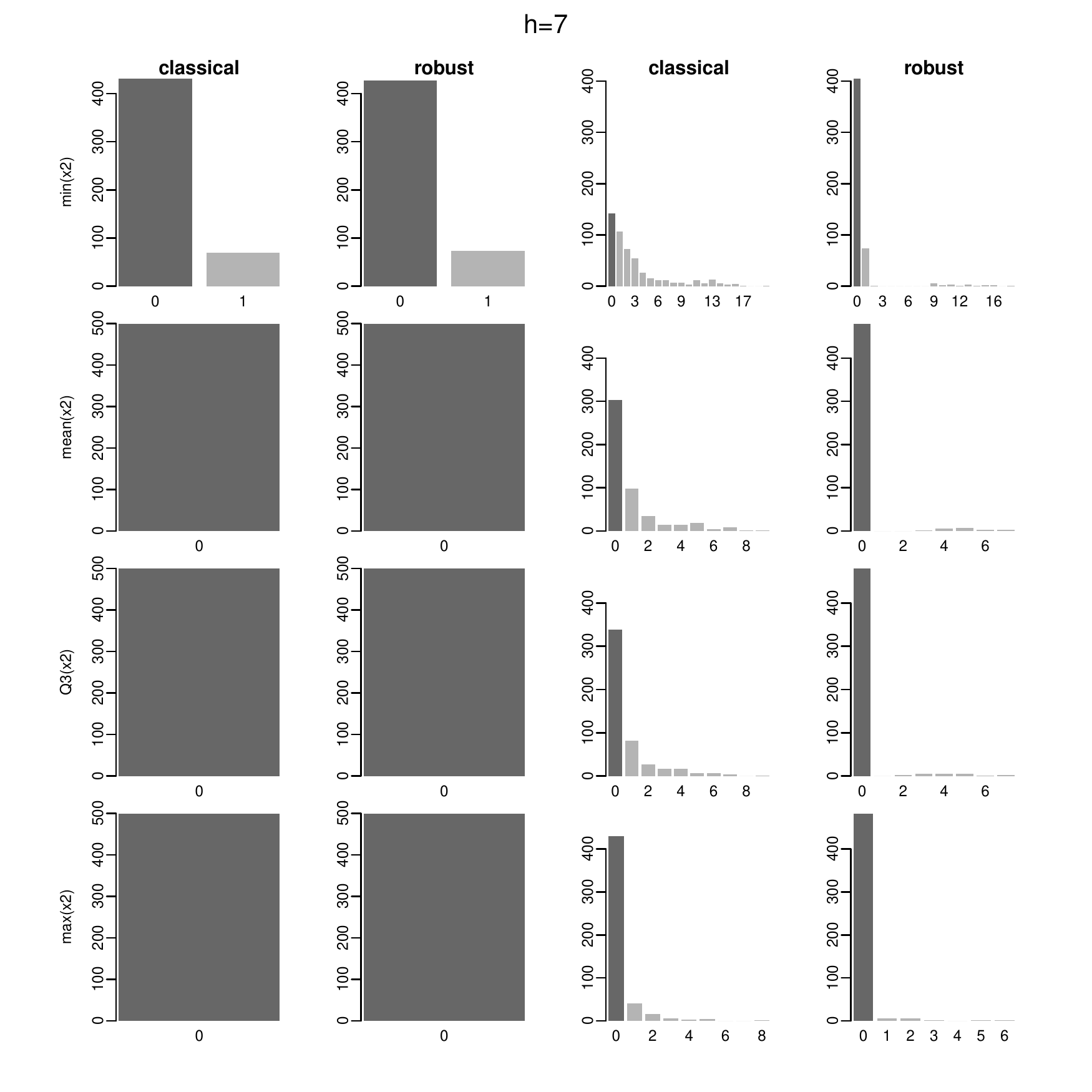}
        \caption{CaRe estimates for $h=7$ for four values of $x_{2}$: its minimum value (first row), its average (second row), its third quartile (third row) and its maximum (fourth row).  The other covariates are fixed at their mean values. The true population value is identified by a darker bar. The first two columns correspond to the clean data setting, whereas the last two columns correspond to the contaminated data setting.}
    \label{fig:CaReDGPDh7x2}
\end{figure}

\begin{figure} [H]
    \centering
    \includegraphics[width=1.1\linewidth]{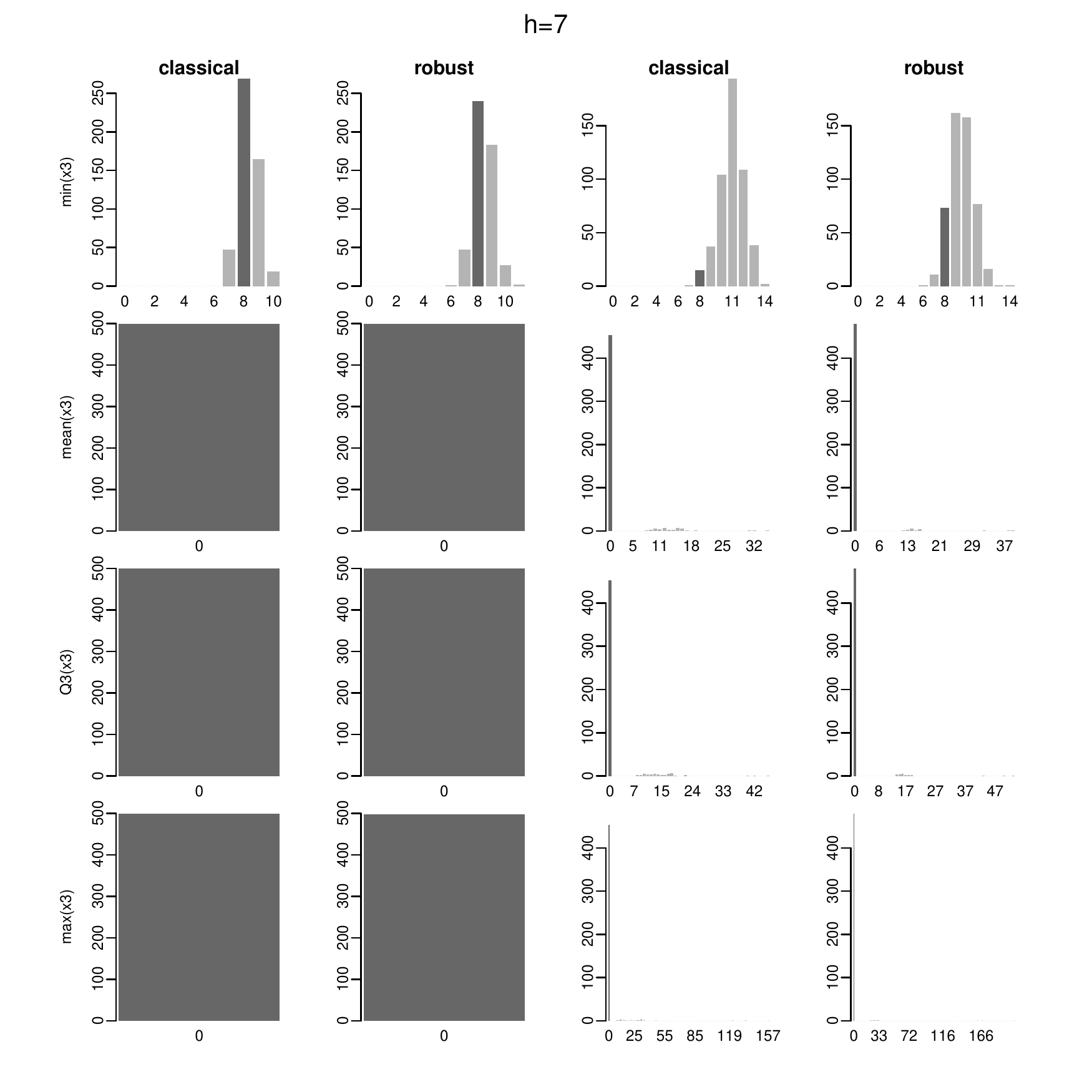}
        \caption{CaRe estimates for $h=7$ for four values of $x_{3}$: its minimum value (first row), its average (second row), its third quartile (third row) and its maximum (fourth row).  The other covariates are fixed at their mean values. The true population value is identified by a darker bar. The first two columns correspond to the clean data setting, whereas the last two columns correspond to the contaminated data setting.}
    \label{fig:CaReDGPDh7x3}
\end{figure}

\begin{figure} [H]
    \centering
    \includegraphics[width=1.1\linewidth]{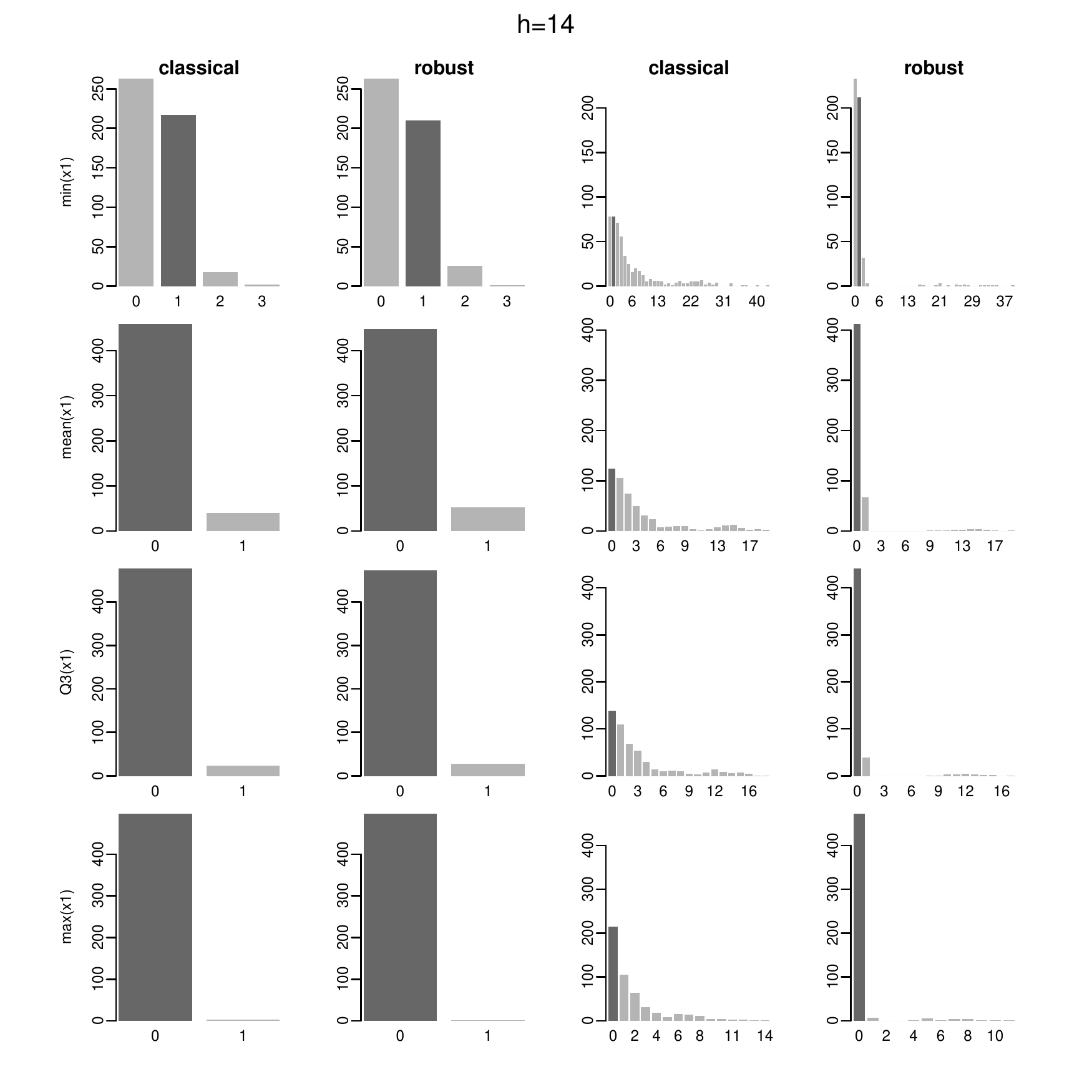}
        \caption{CaRe estimates for $h=14$ for four values of $x_{1}$: its minimum value (first row), its average (second row), its third quartile (third row) and its maximum (fourth row).  The other covariates are fixed at their mean values. The true population value is identified by a darker bar. The first two columns correspond to the clean data setting, whereas the last two columns correspond to the contaminated data setting.}
    \label{fig:CaReDGPDh14x1}
\end{figure}

\begin{figure} [H]
    \centering
    \includegraphics[width=1.1\linewidth]{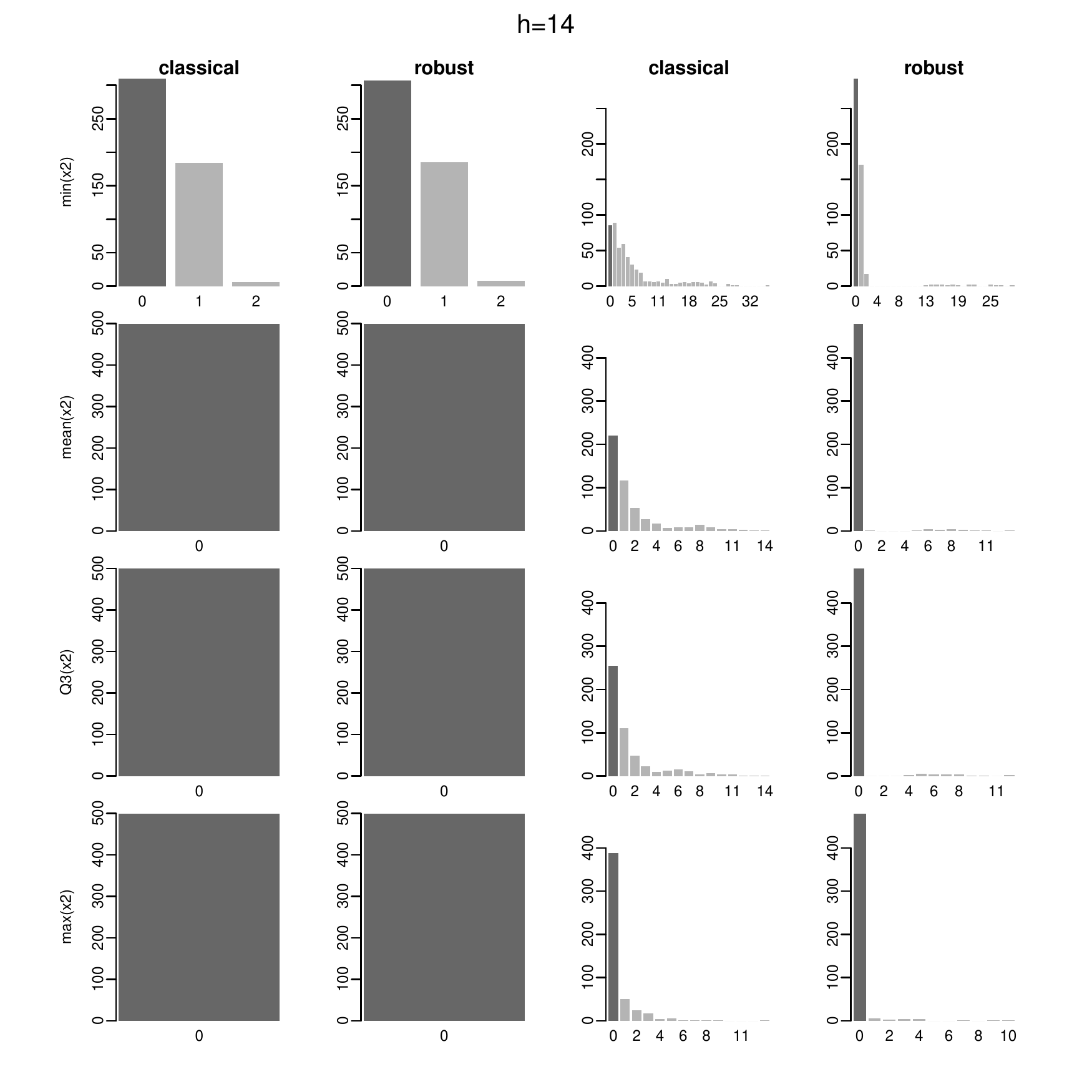}
        \caption{CaRe estimates for $h=14$ for four values of $x_{2}$: its minimum value (first row), its average (second row), its third quartile (third row) and its maximum (fourth row).  The other covariates are fixed at their mean values. The true population value is identified by a darker bar. The first two columns correspond to the clean data setting, whereas the last two columns correspond to the contaminated data setting.}
    \label{fig:CaReDGPDh14x2}
\end{figure}

\begin{figure} [H]
    \centering
    \includegraphics[width=1.1\linewidth]{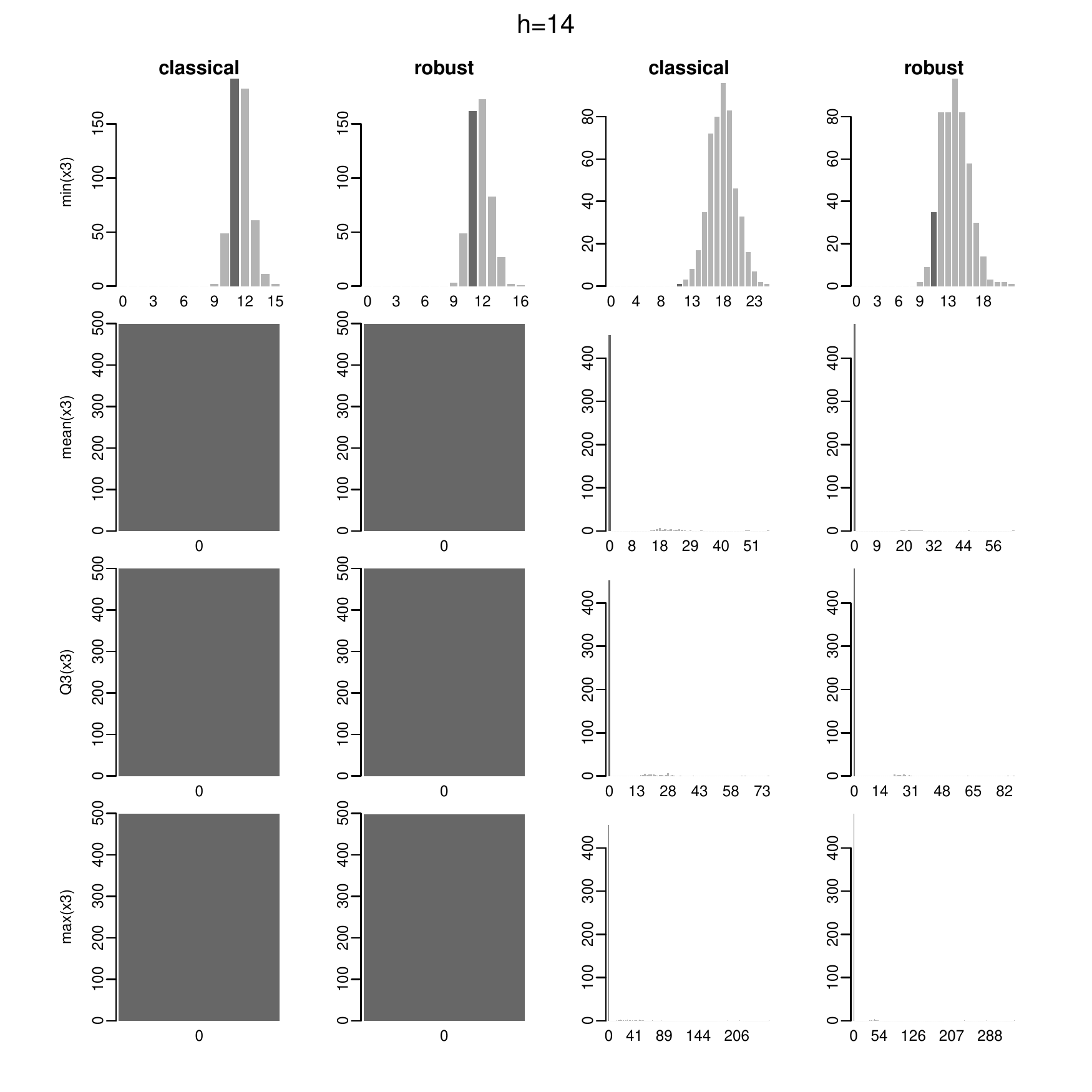}
        \caption{CaRe estimates for $h=14$ for four values of $x_{3}$: its minimum value (first row), its average (second row), its third quartile (third row) and its maximum (fourth row).  The other covariates are fixed at their mean values. The true population value is identified by a darker bar. The first two columns correspond to the clean data setting, whereas the last two columns correspond to the contaminated data setting.}
    \label{fig:CaReDGPDh14x3}
\end{figure}

\begin{figure} [H]
    \centering
    \includegraphics[width=1.1\linewidth]{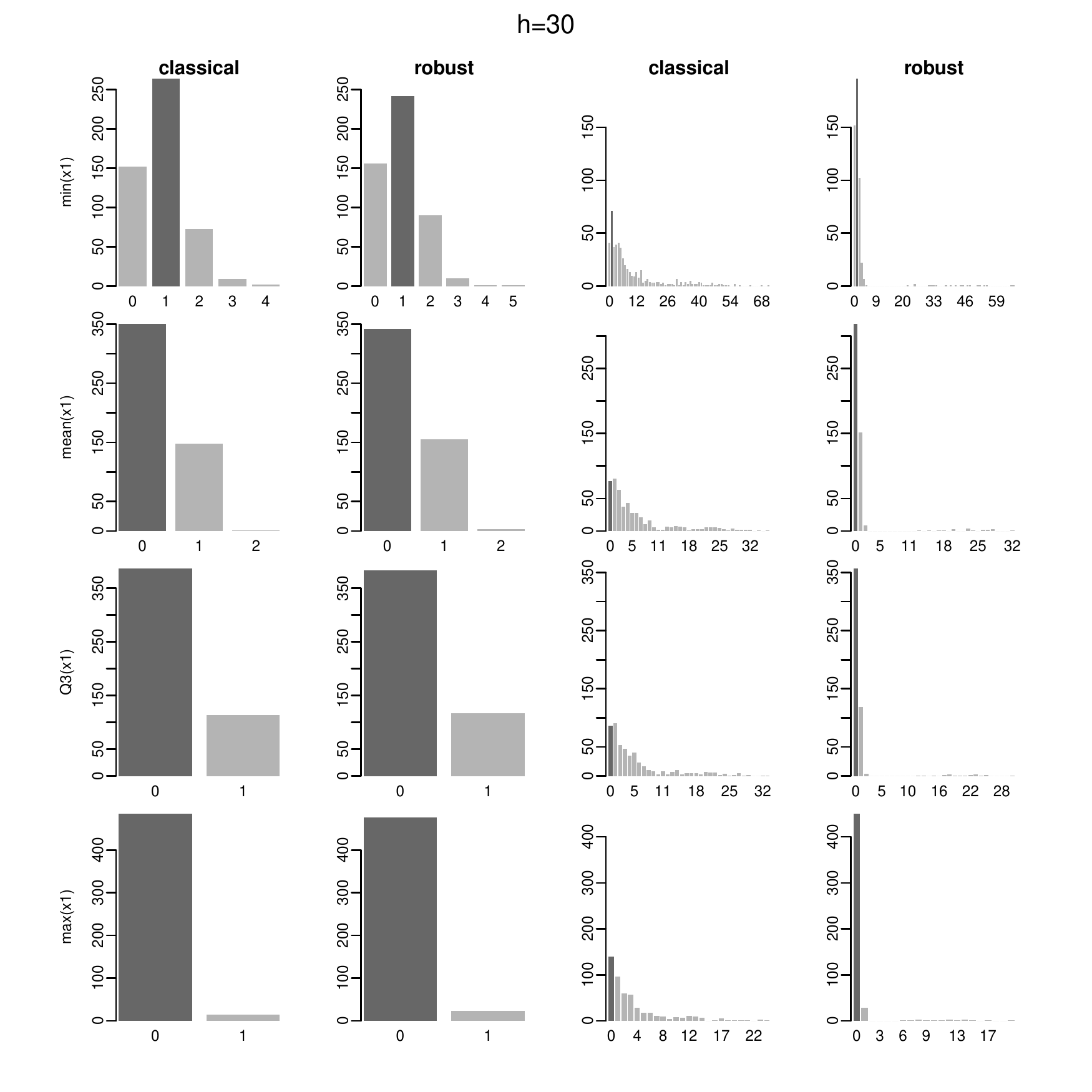}
        \caption{CaRe estimates for $h=30$ for four values of $x_{1}$: its minimum value (first row), its average (second row), its third quartile (third row) and its maximum (fourth row).  The other covariates are fixed at their mean values. The true population value is identified by a darker bar. The first two columns correspond to the clean data setting, whereas the last two columns correspond to the contaminated data setting.}
    \label{fig:CaReDGPDh30x1}
\end{figure}

\begin{figure} [H]
    \centering
    \includegraphics[width=1.1\linewidth]{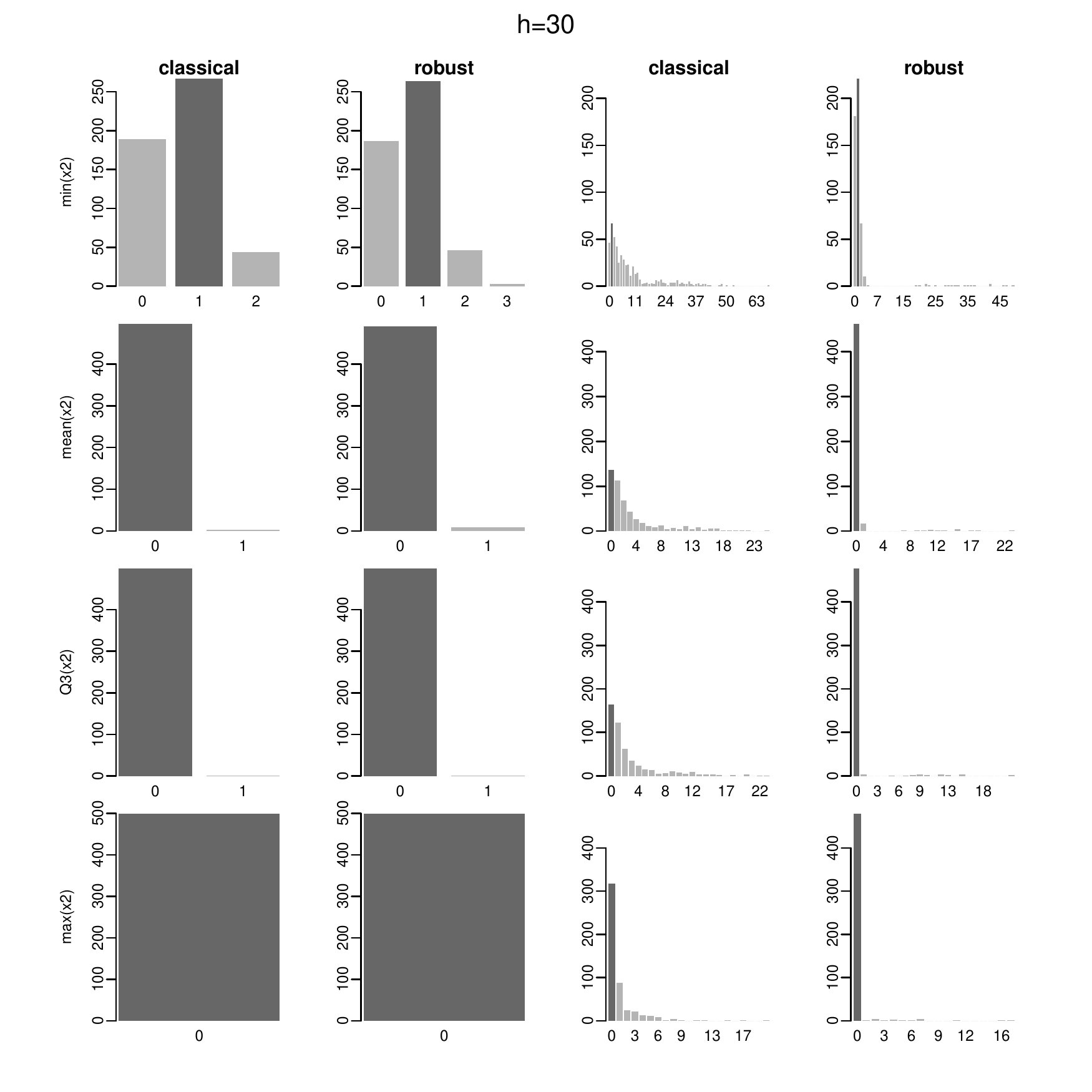}
        \caption{CaRe estimates for $h=30$ for four values of $x_{2}$: its minimum value (first row), its average (second row), its third quartile (third row) and its maximum (fourth row).  The other covariates are fixed at their mean values. The true population value is identified by a darker bar. The first two columns correspond to the clean data setting, whereas the last two columns correspond to the contaminated data setting.}
    \label{fig:CaReDGPDh30x2}
\end{figure}

\begin{figure} [H]
    \centering
    \includegraphics[width=1.1\linewidth]{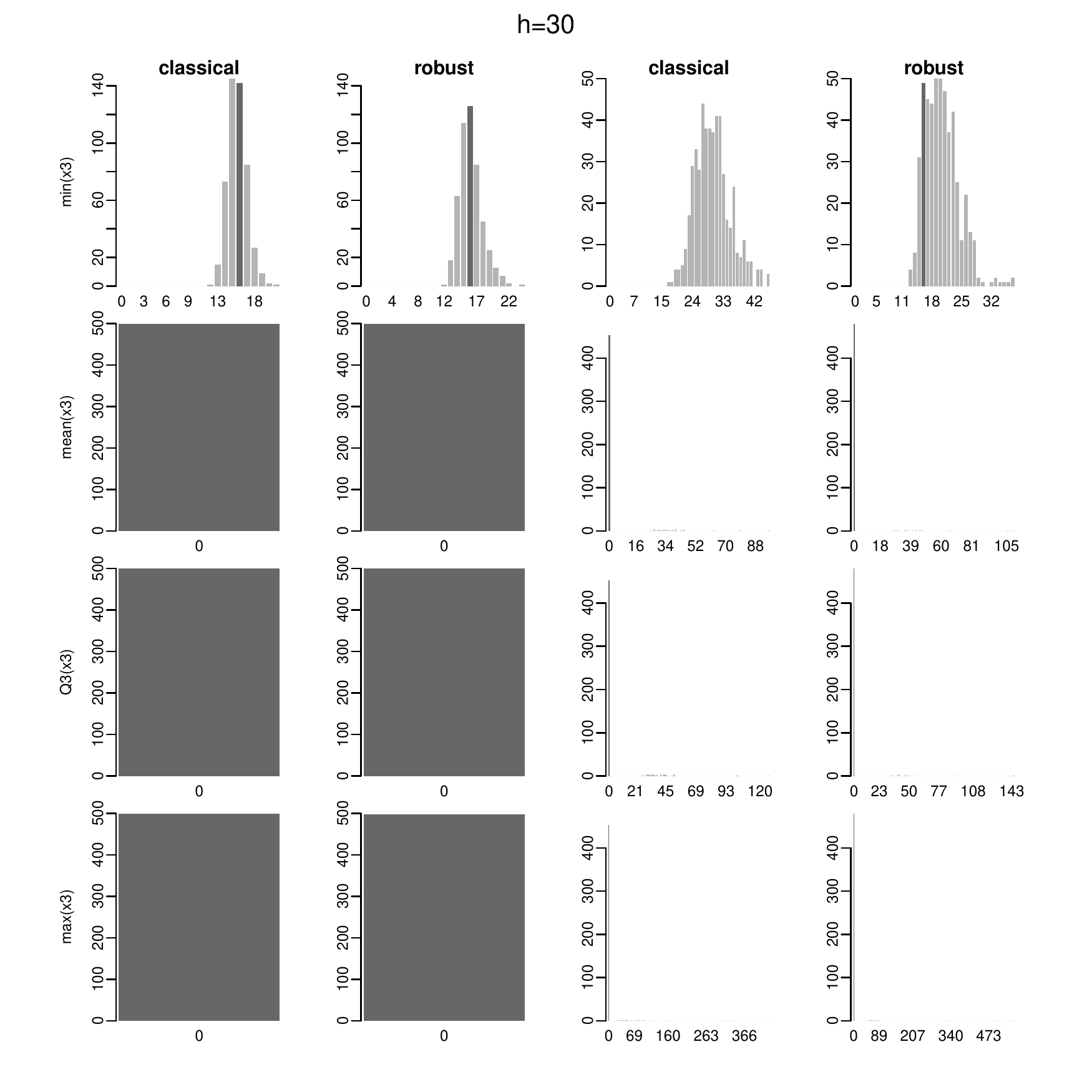}
        \caption{CaRe estimates for $h=30$ for four values of $x_{3}$: its minimum value (first row), its average (second row), its third quartile (third row) and its maximum (fourth row).  The other covariates are fixed at their mean values. The true population value is identified by a darker bar. The first two columns correspond to the clean data setting, whereas the last two columns correspond to the contaminated data setting.}
    \label{fig:CaReDGPDh30x3}
\end{figure}